\documentclass[a4paper,12pt]{article}
\usepackage{expl3}
\usepackage{amsmath,amssymb,mathrsfs,amsthm,tikz,shuffle,paralist}
\usepackage{dsfont}
\usepackage{amsbsy}

\usepackage{xcolor}
\usepackage{relsize}
\usepackage{subfig}

\usepackage[font={small},labelfont=bf]{caption}

\usepackage[compat=1.1.0]{tikz-feynman}
\usetikzlibrary{shapes.misc}
\usetikzlibrary{cd}
\usetikzlibrary{decorations.pathreplacing}
\usepackage{stackengine}
\usepackage[bottom]{footmisc}
\usepackage{booktabs}

\usetikzlibrary{arrows}

\newcommand{\nISP}{{n_{\text{ISP}}}}
\newcommand{\LS}{{\text{LS}}}

\newcommand{\DlogBasis}{\texttt{\textup{DlogBasis}}}

\usepackage{array}
\newcommand{\PreserveBackslash}[1]{\let\temp=\\#1\let\\=\temp}
\newcolumntype{C}[1]{>{\PreserveBackslash\centering}p{#1}}
\newcolumntype{R}[1]{>{\PreserveBackslash\raggedleft}p{#1}}
\newcolumntype{L}[1]{>{\PreserveBackslash\raggedright}p{#1}}

\usepackage{url}

\usepackage{breakurl}
\usepackage[colorlinks=true
,urlcolor=blue
,anchorcolor=blue
,citecolor=blue
,filecolor=blue
,linkcolor=blue
,menucolor=blue
,pagecolor=blue
,linktocpage=true
,pdfproducer=medialab
,pdfa=true
]{hyperref}
\hypersetup{breaklinks=true}

\usepackage[utf8]{inputenc}
\usepackage{graphicx}

\usepackage{jheppub}

\usepackage{etoolbox}
\usepackage{xparse}

\newcommand{\edges}{}
\newcommand{\otheredges}{}

\NewDocumentCommand{\twoloop}{O{0} m m m}{%
\renewcommand{\edges}{}
\ifnum#1=0
	\xappto{\edges}{a -- b[dot], c[dot] -- d,}
\fi
\ifnum#1=1
	\xappto{\edges}{a -- b[dot], c[dot] --[draw=none] d,}
\fi
\ifnum#1=2
	\xappto{\edges}{a --[draw=none] b[dot], c[dot] --[draw=none] d,}
\fi
\ifnum#1=21
	\xappto{\edges}{a --[ghost] b[dot], c[dot] --[draw=none] d,}
\fi
\ifnum#1=10
	\xappto{\edges}{a -- b[dot], c[dot] --[ghost] d,}
\fi
\ifnum#1=210
	\xappto{\edges}{a --[ghost] b[dot], c[dot] --[ghost] d,}
\fi
\ifnum#2=1
	\xappto{\edges}{b --[half right] c,}
\else
	\ifnum#2=5	
		\xappto{\edges}{c -- e2[dot],}
	\else
		\xappto{\edges}{b -- e2[dot],}
	\fi		
  	\foreach \i in {2,...,\numexpr #2\relax} {
  		\pgfmathtruncatemacro{\ip}{\i + 1}
     	\xappto{\edges}{e\i -- f\i,} 
		\ifnum#2=\i     	
     		\ifnum#2=5
     			\xappto{\edges}{e\i -- b,}
     		\else
     			\xappto{\edges}{e\i -- c,}
     		\fi		
     	\else
     		\xappto{\edges}{e\i -- e\ip[dot],} 
		\fi
	}  
\fi
\ifnum#3=1
	\xappto{\edges}{b -- c,}
\else
	\xappto{\edges}{b --[draw=none] c,}
	\xappto{\edges}{b -- g2[dot],}
  	\foreach \i in {2,...,\numexpr #3\relax} {
  		\pgfmathtruncatemacro{\ip}{\i + 1}
     	\xappto{\edges}{g\i -- h\i,} 
		\ifnum#3=\i     	
     		\xappto{\edges}{g\i --[draw=none] k2[dot],}
     	\else
     		\xappto{\edges}{g\i -- g\ip[dot],} 
		\fi
	}  
\fi
\ifnum#4=1
	\ifnum#2=1
		\xappto{\edges}{b --[half left] c,}
	\else
		\ifnum#2=3
			\xappto{\edges}{b --[half left] c,}
		\else
			\xappto{\edges}{b --[half right] c,}
		\fi	
	\fi
\else
	\ifnum#3=1
		\xappto{\edges}{b -- k2[dot],}
	\fi
  	\foreach \i in {2,...,\numexpr #4\relax} {
  		\pgfmathtruncatemacro{\ip}{\i + 1}
     	\xappto{\edges}{k\i -- m\i,} 
		\ifnum#4=\i     	
     		\xappto{\edges}{k\i -- c,}
     	\else
     		\xappto{\edges}{k\i -- k\ip[dot],} 
		\fi
	}  
\fi
\renewcommand{\otheredges}{}
\ifnum#3>1
	\pgfmathtruncatemacro{\ip}{#3}
   	\xappto{\otheredges}{(g\ip) -- (c),} 
   	\xappto{\otheredges}{(b) -- (k2),} 
\fi     	
\edef\mydiagram{%
\noexpand\diagram[small, vertical=b to c,inline=(c),
\ifnum#2<2 
	layered layout 
\fi]{\edges};
\noexpand\diagram*{\otheredges};
}		
\pgfmathparse{(#4 == 1 && #2 !=3) || (#2 == 4 && #3 ==3 && #4==4)|| (#2 == 5 && #3 ==1 && #4==3)|| (#2 == 5 && #3 ==2 && #4==2)|| (#2 == 5 && #3 ==2 && #4==4)|| (#2 == 5 && #3 ==3 && #4==3)|| (#2 == 2 && #3 ==2 && #4==2)|| (#2 == 4 && #3 ==1 && #4==3)|| (#2 == 4 && #3 ==2 && #4==2)|| (#2 == 4 && #3 ==2 && #4==3)|| (#2 == 3 && #3 ==3 && #4==3) ? -1 : 1}%
  \let\xsfactor\pgfmathresult
\begin{tikzpicture}[scale=0.5,xscale=\xsfactor,baseline=($(b)!0.7!(c)$)
]
\begin{feynman}
\mydiagram
\end{feynman}
\end{tikzpicture}
}

\usetikzlibrary{external}
\usepackage[textsize=tiny]{todonotes}  
 
\makeatletter
\renewcommand{\todo}[2][]{\tikzexternaldisable\@todo[#1]{#2}\tikzexternalenable}
\makeatother

\begin{document}

\title{The spectrum of Feynman-integral geometries at two loops}

\author[a]{Piotr Bargie\l{}a,}
\emailAdd{pbargiel@ed.ac.uk}
\author[b]{Hjalte Frellesvig,}
\emailAdd{0025056@zju.edu.cn}
\author[c]{Robin Marzucca,}
\emailAdd{robin.marzucca@physik.uzh.ch}
\author[d,e]{Roger Morales,}
\emailAdd{rmespasa@umich.edu}
\author[f]{Florian Seefeld,}
\emailAdd{fseefeld@imada.sdu.dk}
\author[e,f]{Matthias Wilhelm,}
\emailAdd{mwilhelm@imada.sdu.dk}
\author[c,g,h]{and Tong-Zhi Yang}
\emailAdd{tongzhi.yang@m.scnu.edu.cn}

\affiliation[a]{%
Higgs Centre for Theoretical Physics, School of Physics and Astronomy, The University of Edinburgh, Edinburgh EH9 3FD, Scotland, UK}
\affiliation[b]{%
Zhejiang Institute of Modern Physics, School of Physics, Zhejiang University, Hangzhou 310027, China}
\affiliation[c]{%
Physik-Institut, Universit\"at Z\"urich, Winterthurerstrasse 190, 8057 Z\"urich, Switzerland}
\affiliation[d]{%
Leinweber Institute for Theoretical Physics, Randall Laboratory of Physics, University of Michigan, 450 Church St, Ann Arbor, MI 48109-1040, USA}
\affiliation[e]{%
Niels Bohr International Academy, Niels Bohr Institute, Copenhagen University, 2100 Copenhagen \O{}, Denmark}
\affiliation[f]{%
Center for Quantum Mathematics, Department of Mathematics and Computer Science, University of Southern Denmark, 5230 Odense M, Denmark}

\affiliation[g]{State Key Laboratory of Nuclear Physics and Technology, Institute of Quantum Matter, South China Normal University, Guangzhou 510006, China}
\affiliation[h]{%
Guangdong Basic Research Center of Excellence for Structure and Fundamental Interactions of Matter, Guangdong Provincial Key Laboratory of Nuclear Science, Guangzhou 510006, China
}

\preprint{LITP-25-15, ZU-TH 81/25}

\date{\today}%

\abstract{%
We provide a complete classification of the Feynman-integral geometries at two-loop order in four-dimensional Quantum Field Theory with standard quadratic propagators. Concretely, we consider a finite basis of integrals in the 't Hooft--Veltman scheme, i.e.\ with $D$-dimensional loop momenta and four-dimensional external momenta, which belong to 79 independent topologies, or sectors. Then, we analyze the leading singularities of the integrals in those sectors for generic values of the masses and momenta, using the loop-by-loop Baikov representation. Aside from the Riemann sphere, we find that elliptic curves, hyperelliptic curves of genus 2 and 3 as well as K3 surfaces occur. Moreover, we find a smooth and non-degenerate Del Pezzo surface of degree 2, a particular Fano variety known to be rationalizable, resulting in a curve of geometric genus 3. These geometries determine the space of functions relevant for Quantum Field Theories at two-loop order, including in the Standard Model.
\enlargethispage{\baselineskip}
}

\maketitle

\section{Introduction}
\label{sec:intro}
 
Feynman integrals are a key ingredient for precision predictions within Quantum Field Theory (QFT). Specifically, evaluating Feynman integrals is an essential step for calculating scattering amplitudes, which are used to obtain the physical observables that can be compared to experiments. With the upcoming high-luminosity upgrade to the Large Hadron Collider (LHC), theoretical predictions for numerous further processes are required at higher precision than currently available~\cite{Andersen:2024czj,Huss:2025nlt}, including many processes still at two-loop order. 

In four dimensions, all Feynman integrals at one-loop order can be expressed in terms of multiple polylogarithms,\footnote{This has also been conjectured to hold for all one-loop Feynman integrals in general dimensions; see ref.~\cite{Papathanasiou:2025stn}.} which are iterated integrals on the Riemann sphere that are by now well understood~\cite{Chen:1977oja,Goncharov:1995ifj}.
However, at higher loop orders other, more complicated functions can occur, which stem from integrals over more intricate geometries; see ref.~\cite{Bourjaily:2022bwx} for a recent review. During the last few years, Feynman integrals involving elliptic curves~\cite{Sabry:1962rge,Broadhurst:1993mw,Laporta:2004rb,Caron-Huot:2012awx,Adams:2013nia,Bloch:2013tra,Adams:2014vja,Remiddi:2016gno,Adams:2016xah,Broedel:2017siw,Kristensson:2021ani,Giroux:2022wav,Morales:2022csr,McLeod:2023qdf,Stawinski:2023qtw,Giroux:2024yxu,Spiering:2024sea}, hyperelliptic curves \cite{Huang:2013kh,Marzucca:2023gto,Duhr:2024uid} as well as Calabi--Yau (CY) geometries of arbitrarily high dimension~\cite{Primo:2017ipr,Bourjaily:2018ycu,Bourjaily:2018yfy,Bonisch:2021yfw,Broedel:2021zij,Duhr:2022pch,Lairez:2022zkj,Pogel:2022vat,Duhr:2022dxb,Cao:2023tpx,Doran:2023yzu,McLeod:2023doa,Duhr:2023eld,Frellesvig:2023bbf,Klemm:2024wtd,Duhr:2024hjf,Frellesvig:2024zph,Frellesvig:2024rea,Duhr:2025ppd,Duhr:2025lbz,Maggio:2025jel,Brammer:2025rqo,e-collaboration:2025frv,Duhr:2025kkq,Pogel:2025bca,Duhr:2025xyy} have been identified, with relevance to the Standard Model~\cite{Sabry:1962rge,Broadhurst:1993mw,Adams:2018bsn,Adams:2018kez,Broedel:2019hyg,Abreu:2019fgk,Duhr:2024bzt,Forner:2024ojj,Marzucca:2025eak,Becchetti:2025rrz,Bargiela:2025nqc,Coro:2025vgn}, classical gravity~\cite{Bern:2021dqo,Dlapa:2021npj,Dlapa:2022wdu,Jakobsen:2023ndj,Frellesvig:2023bbf,Klemm:2024wtd,Driesse:2024xad,Frellesvig:2024zph,Frellesvig:2024rea,Brammer:2025rqo,Driesse:2024feo} as well as supersymmetric theories~\cite{Caron-Huot:2012awx,Bourjaily:2018ycu,Bourjaily:2018yfy,Kristensson:2021ani,Morales:2022csr,Duhr:2022pch,Duhr:2023eld,Cao:2023tpx,McLeod:2023qdf,McLeod:2023doa,Duhr:2024hjf,Bern:2024adl}. However, a complete understanding of which geometries and functions occur at high-loop orders, including at two-loop order, is still missing.

In this paper, we close this gap by classifying all geometries that can occur at two-loop order in four-dimensional Quantum Field Theory with standard quadratic propagators, which thus determines the corresponding function space, including in the theory of Quantum Chromodynamics (QCD) and the Standard Model. Specifically, we base our classification on a recently constructed finite basis of two-loop integrals~\cite{Bargiela:2024rul,Bargiela:2025nqc} in the 't Hooft--Veltman scheme~\cite{tHooft:1972tcz} (see also ref.~\cite{Mastrolia:2016dhn}). In this scheme, the loop momenta are in $D=4-2\varepsilon$ dimensions but the external momenta are kept in $D=4$. The integrals in the basis belong to 79 independent sectors, or integral topologies. To achieve our classification, we use a set of techniques recently developed for Feynman integrals in the context of classical gravity\ \cite{Frellesvig:2023bbf,Frellesvig:2024zph,Brammer:2025rqo}. In particular, we use the loop-by-loop Baikov representation \cite{Frellesvig:2017aai,Frellesvig:2024ymq} to analyze the leading singularities \cite{Cachazo:2008vp,Arkani-Hamed:2010pyv} of the integrals in these sectors, which characterizes their associated geometry at the level of the maximal cut and thus the function space. We find that simple properties of the Feynman-integral sector, such as the number of external legs in each individual loop, suffice to bound the dimension and complexity of the geometries. This bound is actually saturated in surprisingly many integral sectors, resulting in only a few special cases that are analyzed separately.

\begin{figure}[t]
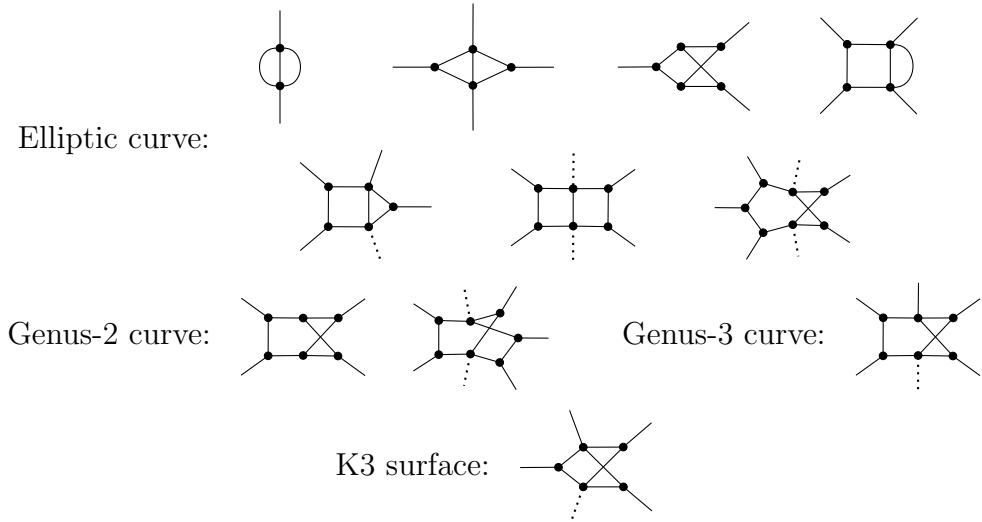

\begin{center}
\hspace{2.3cm} $\twoloop[0]{1}{1}{1} \qquad \hspace{0.3cm} \twoloop[0]{2}{1}{2} \qquad \twoloop[2]{2}{2}{2} \quad \hspace{0.5cm} \twoloop[0]{3}{1}{1}$ \\[-0.1cm]
\hspace{-10cm}Elliptic curve: \\[-0.1cm]
\hspace{2cm} $\twoloop[10]{3}{1}{2} \qquad \enspace \twoloop[210]{3}{1}{3} \enspace \qquad \twoloop[210]{4}{2}{2}$  
\\[0.2cm]
Genus-2 curve: $\enspace \twoloop[2]{3}{2}{2} \enspace \quad \twoloop[210]{3}{2}{3}$
\qquad
Genus-3 curve: $\enspace \twoloop[10]{3}{2}{2} $
\\[0.2cm]
K3 surface: $\enspace \twoloop[10]{2}{2}{2}$
\end{center}
\vspace{-0.5cm}
    \caption{Independent two-loop Feynman-integral topologies in the 't Hooft--Veltman scheme that contain non-trivial geometries at the level of the maximal cut, and which contribute to order $\mathcal{O}(\varepsilon^0)$. We consider all internal propagators to be massive, all external legs off-shell, and take generic values of the masses and momenta. External legs that do not need to be present are dotted.}
    \label{fig: non-trivial_diagrams_2loop}
\end{figure}

\begin{figure}[t]
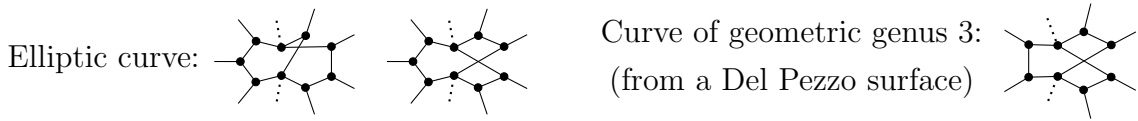

\begin{center}
Elliptic curve: $\twoloop[210]{4}{2}{4} \quad \twoloop[210]{4}{3}{3}$ \qquad
$\substack{\hspace{0.0cm} \text{\normalsize{Curve of geometric genus 3:}}\\[0.2cm] \text{\normalsize{(from a Del Pezzo surface)}}} \ $ $\twoloop[210]{3}{3}{3}$
\end{center}
\vspace{-0.5cm}
    \caption{Independent two-loop Feynman-integral topologies in the 't Hooft--Veltman scheme that contain non-trivial geometries at the level of the maximal cut, and which are evanescent, i.e.\ only contribute to order $\mathcal{O}(\varepsilon^1)$ or higher. We consider all internal propagators to be massive, all external legs off-shell, and take generic values of the masses and momenta. External legs that do not need to be present are dotted.}
    \label{fig: non-trivial_diagrams_2loop_evanescent}
\end{figure}

In total, we find that at most elliptic curves occur in planar two-loop Feynman integrals, whereas non-planar two-loop Feynman integrals can in addition contain hyperelliptic curves of genus 2 and 3 as well as K3 surfaces, i.e.\ CY geometries of dimension two. Moreover, we observe the appearance of a smooth and non-degenerate Del Pezzo surface of degree 2, a particular type of Fano variety that is known to be rationalizable~\cite{Schicho2005}.\footnote{See e.g.\ refs.~\cite{Bloch:2016izu,Schimmrigk:2024xid,delaCruz:2025szs} for a discussion of Fano varieties related to Feynman integrals.} After the rationalization, the resulting geometry is a curve of geometric genus 3, which is not necessarily hyperelliptic. A complete list of the two-loop Feynman-integral topologies that contain non-trivial geometries and contribute to order $\mathcal{O}(\varepsilon^{0})$ is given in fig.~\ref{fig: non-trivial_diagrams_2loop}. Similarly, a list of integral topologies that depend on non-trivial geometries but are evanescent, i.e.\ whose independent contributions only begin at order $\mathcal{O}(\varepsilon^{1})$ or higher, is given in fig.~\ref{fig: non-trivial_diagrams_2loop_evanescent}. 
 
While a subset of these integrals were already known to depend on non-trivial geometries, they were mostly studied in certain massless or equal-mass limits. Instead, the classification presented in this paper is completely general, and we argue that there exists a notion of a unique geometry associated to each integral topology. Still, simplifications are expected to occur in special kinematics, such as a massless limit, where the geometries can degenerate. Let us also note that in ref.~\cite{Doran:2023yzu} a large-scale analysis of the geometries of two-loop Feynman integrals was performed, notably finding an upper bound on the complexity of the geometry of planar integrals; see also ref.~\cite{Lairez:2022zkj} for several specific examples.

The remainder of this paper is structured as follows. In sec.~\ref{sec:intro_basis_integrals}, we review the 't~Hooft--Veltman scheme and introduce the 79 independent integral topologies that need to be considered for the purpose of classifying the geometries arising at two-loop order. Then, in sec.~\ref{sec:classification}, we review the techniques that can be used to identify the occurring geometries, specializing in the analysis of leading singularities via the loop-by-loop Baikov representation. As a further tool to identify these geometries, we use the notion of complete intersection manifolds and configuration matrices~\cite{Green:1986ck,Candelas:1987kf,Hubsch:1992nu}. Thereafter, we present a general analysis of the possible geometries that can occur at two loops, based solely on the result of the maximal cut of the corresponding integrals. Then, in sec.~\ref{sec:special_cases}, we discuss several special cases where the analysis is more subtle. Lastly, in sec.~\ref{sec:conclusions}, we present our conclusions and discuss further research directions. 

This paper also includes three appendices. First, in app.~\ref{app:factorization_PF}, we consider the factorization of Picard--Fuchs operators for multi-scale Feynman integrals, which can serve as an alternative means of studying the geometries in the integrals. Specifically, we demonstrate using an example that this factorization can be quite non-trivial for multi-scale integrals even in cases admitting a $d$log form at the maximal cut. Then, in app.~\ref{app:gram}, we investigate the general structure and the degree of the Baikov polynomials from the associated Gram determinants. Finally, in app.~\ref{app:rationalization_Del_Pezzo}, we present the details for the rationalization of a generic Del Pezzo surface of degree 2. As ancillary files, we include \texttt{Mathematica} notebooks detailing the specific parametrization, loop-by-loop Baikov representation and the full analysis of leading singularities for the integrals in the 79 independent integral topologies.

\textbf{Note added:} While this paper was under preparation, the work \cite{delaCruz:2025szs} appeared on the arXiv, which has some overlap with regards to the Del Pezzo surface.

\section{Feynman integrals in the 't Hooft--Veltman scheme}
\label{sec:intro_basis_integrals}

In this section, we explain which two-loop integral topologies have to be considered in order to span all of the function space in four-dimensional Quantum Field Theory with standard quadratic propagators. Moreover, we briefly review their kinematic dependence. 

Since generic Feynman integrals can be singular in four dimensions, we dimensionally regulate them by integrating the loop momenta in $D=4-2\varepsilon$ dimensions.\footnote{In some cases, we find it more convenient to work in $D=2-2\varepsilon$ or $D=6-2\varepsilon$ dimensions, depending on the number of external legs. Note that, since the corresponding results can be related to the four-dimensional ones via dimension-shift identities~\cite{Tarasov:1996br,Lee:2012te}, the Feynman-integral geometry remains the same.} In the Conventional Dimensional Regularization scheme~\cite{Bollini:1972ui,tHooft:1972tcz}, where the external legs are also treated as $D$-dimensional, there is no upper bound on the number of independent integral topologies appearing at a fixed loop order. By contrast, as recently pointed out in refs.~\cite{Bargiela:2024rul,Bargiela:2025nqc}, keeping the external legs purely in four dimensions, as prescribed by the 't Hooft--Veltman scheme~\cite{tHooft:1972tcz}, allows us to define a finite basis of Feynman integrals
 for two-loop scattering amplitudes. Thus, for the rest of this paper, we will consider all integrals within the 't~Hooft--Veltman scheme.
Let us note that the choice of scheme cannot change observables and other scheme-independent quantities, such that we are free to pick this convenient scheme for our analysis.

\subsection{Independent two-loop integral topologies}
\label{subsec:integral_basis}

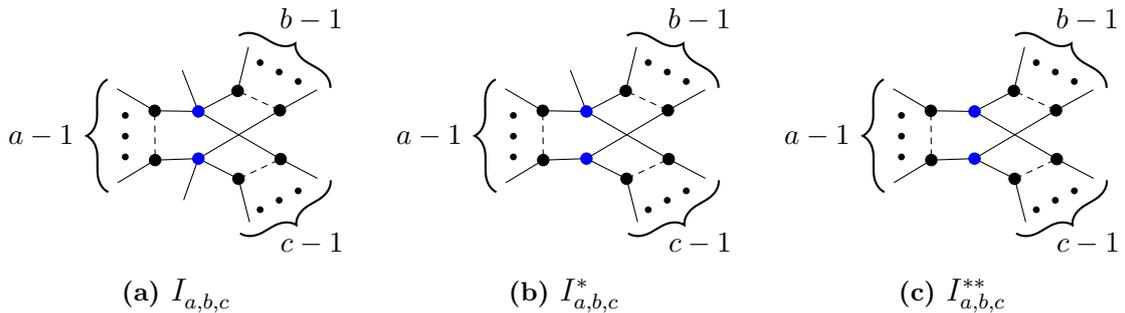
\begin{figure}[t]
\centering
\subfloat[\normalsize $I^{}_{a,b,c}$]{\begin{tikzpicture}[scale=0.5,xscale=1,baseline=($(b)!0.5!(c)$)
]
\begin{feynman}
\diagram[vertical= b to c]{
a -- b[dot, blue], c[dot, blue] -- d,
b -- e1[dot],
e1 --[dash pattern=on 2.5pt off 2pt] e2[dot],
e1 -- f1,
e2 -- f2,
e2 -- c,
b --[draw=none] c,
b -- g1[dot],
g1 --[dash pattern=on 2.5pt off 2pt] g2[dot],
g1 -- h1,
g2 -- h2,
g2 --[draw=none] k1[dot],
k1 --[dash pattern=on 2.5pt off 2pt] k2[dot],
k1 -- m1,
k2 -- m2,
k2 -- c,
};
\diagram*{
(g2) -- (c),
(b) -- (k1),
};
\draw [decorate,decoration={brace,amplitude=8pt}, thick]
  ($(h1) + (-0.25,0.3)$) -- ($(h2) + (0.35,0.05)$)
  node[midway, xshift=10pt, yshift=16pt] {\small $b-1$};
\draw [decorate,decoration={brace,amplitude=8pt}, thick]
  ($(m1) + (0.35,-0.05)$) -- ($(m2) + (-0.25,-0.3)$)
  node[midway, xshift=10pt, yshift=-14pt] {\small $c-1$};
  \draw [decorate,decoration={brace,amplitude=8pt}, thick]
  ($(f2) + (-0.25,-0.25)$) -- ($(f1) + (-0.25,0.25)$)
  node[midway, xshift=-25pt, yshift=0pt] {\small $a-1$};
  \node at (-1.5,-1.2) [circle,fill,inner sep=0.9pt]{};
  \node at (-1.5,-1.75) [circle,fill,inner sep=0.9pt]{};
  \node at (-1.5,-2.3) [circle,fill,inner sep=0.9pt]{};
  \node at (2.05,0.2) [circle,fill,inner sep=0.9pt]{};
  \node at (2.55,-0.05) [circle,fill,inner sep=0.9pt]{};
  \node at (3.05,-0.3) [circle,fill,inner sep=0.9pt]{};
  \node at (2.05,-3.7) [circle,fill,inner sep=0.9pt]{};
  \node at (2.55,-3.45) [circle,fill,inner sep=0.9pt]{};
  \node at (3.05,-3.2) [circle,fill,inner sep=0.9pt]{};
\end{feynman}
\end{tikzpicture}} \quad \subfloat[\normalsize $I^{*}_{a,b,c}$]{\begin{tikzpicture}[scale=0.5,xscale=1,baseline=($(b)!0.5!(c)$)
]
\begin{feynman}
\diagram[vertical= b to c]{
a -- b[dot, blue], c[dot, blue] --[draw=none] d,
b -- e1[dot],
e1 --[dash pattern=on 2.5pt off 2pt] e2[dot],
e1 -- f1,
e2 -- f2,
e2 -- c,
b --[draw=none] c,
b -- g1[dot],
g1 --[dash pattern=on 2.5pt off 2pt] g2[dot],
g1 -- h1,
g2 -- h2,
g2 --[draw=none] k1[dot],
k1 --[dash pattern=on 2.5pt off 2pt] k2[dot],
k1 -- m1,
k2 -- m2,
k2 -- c,
};
\diagram*{
(g2) -- (c),
(b) -- (k1),
};
\draw [decorate,decoration={brace,amplitude=8pt}, thick]
  ($(h1) + (-0.25,0.3)$) -- ($(h2) + (0.35,0.05)$)
  node[midway, xshift=10pt, yshift=16pt] {\small $b-1$};
\draw [decorate,decoration={brace,amplitude=8pt}, thick]
  ($(m1) + (0.35,-0.05)$) -- ($(m2) + (-0.25,-0.3)$)
  node[midway, xshift=10pt, yshift=-14pt] {\small $c-1$};
  \draw [decorate,decoration={brace,amplitude=8pt}, thick]
  ($(f2) + (-0.25,-0.25)$) -- ($(f1) + (-0.25,0.25)$)
  node[midway, xshift=-25pt, yshift=0pt] {\small $a-1$};
  \node at (-1.5,-1.2) [circle,fill,inner sep=0.9pt]{};
  \node at (-1.5,-1.75) [circle,fill,inner sep=0.9pt]{};
  \node at (-1.5,-2.3) [circle,fill,inner sep=0.9pt]{};
  \node at (2.05,0.2) [circle,fill,inner sep=0.9pt]{};
  \node at (2.55,-0.05) [circle,fill,inner sep=0.9pt]{};
  \node at (3.05,-0.3) [circle,fill,inner sep=0.9pt]{};
  \node at (2.05,-3.7) [circle,fill,inner sep=0.9pt]{};
  \node at (2.55,-3.45) [circle,fill,inner sep=0.9pt]{};
  \node at (3.05,-3.2) [circle,fill,inner sep=0.9pt]{};
\end{feynman}
\end{tikzpicture}} \quad \subfloat[\normalsize $I^{**}_{a,b,c}$]{\begin{tikzpicture}[scale=0.5,xscale=1,baseline=($(b)!0.5!(c)$)
]
\begin{feynman}
\diagram[vertical= b to c]{
a --[draw=none] b[dot, blue], c[dot, blue] --[draw=none] d,
b -- e1[dot],
e1 --[dash pattern=on 2.5pt off 2pt] e2[dot],
e1 -- f1,
e2 -- f2,
e2 -- c,
b --[draw=none] c,
b -- g1[dot],
g1 --[dash pattern=on 2.5pt off 2pt] g2[dot],
g1 -- h1,
g2 -- h2,
g2 --[draw=none] k1[dot],
k1 --[dash pattern=on 2.5pt off 2pt] k2[dot],
k1 -- m1,
k2 -- m2,
k2 -- c,
};
\diagram*{
(g2) -- (c),
(b) -- (k1),
};
\draw [decorate,decoration={brace,amplitude=8pt}, thick]
  ($(h1) + (-0.25,0.3)$) -- ($(h2) + (0.35,0.05)$)
  node[midway, xshift=10pt, yshift=16pt] {\small $b-1$};
\draw [decorate,decoration={brace,amplitude=8pt}, thick]
  ($(m1) + (0.35,-0.05)$) -- ($(m2) + (-0.25,-0.3)$)
  node[midway, xshift=10pt, yshift=-14pt] {\small $c-1$};
  \draw [decorate,decoration={brace,amplitude=8pt}, thick]
  ($(f2) + (-0.25,-0.25)$) -- ($(f1) + (-0.25,0.25)$)
  node[midway, xshift=-25pt, yshift=0pt] {\small $a-1$};
  \node at (-1.5,-1.2) [circle,fill,inner sep=0.9pt]{};
  \node at (-1.5,-1.75) [circle,fill,inner sep=0.9pt]{};
  \node at (-1.5,-2.3) [circle,fill,inner sep=0.9pt]{};
  \node at (2.05,0.2) [circle,fill,inner sep=0.9pt]{};
  \node at (2.55,-0.05) [circle,fill,inner sep=0.9pt]{};
  \node at (3.05,-0.3) [circle,fill,inner sep=0.9pt]{};
  \node at (2.05,-3.7) [circle,fill,inner sep=0.9pt]{};
  \node at (2.55,-3.45) [circle,fill,inner sep=0.9pt]{};
  \node at (3.05,-3.2) [circle,fill,inner sep=0.9pt]{};
\end{feynman}
\end{tikzpicture}}
\caption{Graphical representation of our notation for two-loop Feynman-integral topologies, which feature two vertices (highlighted in blue) where three propagators meet. These vertices can be connected via three distinct paths, containing $a$, $b$ and $c$ propagators, respectively. We denote the respective integral topology by $I^{}_{a,b,c}$ when both cubic vertices are attached to external legs (\textbf{a}), by $I^{*}_{a,b,c}$ when only one of the two vertices has external legs (\textbf{b}), and by $I^{**}_{a,b,c}$ when neither are connected to external legs (\textbf{c}).}
\label{fig:integral_notation}
\end{figure}

To begin with, we only need to consider Feynman integrals that are not products of one-loop integrals, since those admit a $d$log form and are already understood. For the remaining two-loop Feynman-integral topologies, we use the following notation, which combines the notations of refs.~\cite{Doran:2023yzu} and~\cite{Gluza:2010ws}. Specifically, all corresponding two-loop graphs contain two vertices where three internal propagators meet, and three paths connecting these vertices; see fig.~\ref{fig:integral_notation}. Let $a,b,c$ denote the number of propagators along these paths. Since the arrangement of the paths is arbitrary, we choose $a\geq c\geq b \geq 1$ without loss of generality.\footnote{If $b=0$, the two-loop integral factorizes into a product of one-loop integrals.} We denote such an integral topology by $I_{a,b,c}$ if both vertices contain external legs, by $I^*_{a,b,c}$ if one of them contains an external leg, and by $I^{**}_{a,b,c}$ if neither of them contain an external leg; see fig.~\ref{fig:integral_notation} for a graphical representation. Lastly, a topology is planar if and only if $b=1$.

In the 't Hooft--Veltman scheme, where the external momenta $p_i^\mu$ are constrained to four dimensions, at most four of those momenta can be linearly independent. This means that at most $11$ independent scalar products involving the loop momenta $k_j^\mu$ can be formed, i.e.\ $k_1^2$, $k_2^2$, $k_1 \cdot k_2$, and 8 of the form $p_i \cdot k_j$. As a consequence, any integral with more than $11$ propagators can be expressed in terms of integrals with fewer propagators using partial fraction identities; see ref.~\cite{Bargiela:2024rul} for a closed partial fraction formula to achieve this. In fact, we can apply the same argument to the individual paths $a,b,c$. For any given path there exists a parametrization of the integral such that at most $5$ independent scalar products involving the corresponding loop momentum $k^\mu$ appear: 4 scalar products $k \cdot p_j$ along with $k^2$. Therefore, any path with more than $5$ propagators can be reduced via partial fraction identities to a linear combination of paths with fewer propagators. Overall, we thus have the constraints
\begin{equation}
5 \geq a \geq c \geq b \geq 1 \,, \qquad \text{and} \qquad\; 11 \geq a + b + c\,.
\end{equation}
This results in $28$ possibilities for each $I_{a,b,c}$, $I^*_{a,b,c}$ and $I^{**}_{a,b,c}$, yielding a total of $84$ integral topologies to be studied, in accordance with ref.~\cite{Bargiela:2025nqc}. Of those, the integral topologies containing more than 8 propagators yield evanescent contributions~\cite{Bargiela:2025nqc}, i.e.\ they only contribute at order $\mathcal{O}(\varepsilon^1)$ or higher at two loops. We will nonetheless include them in our analysis of geometries for completeness, and since $\mathcal{O}(\varepsilon^1)$ contributions from the two-loop integrals are required at higher loop orders.

We can, however, exclude a few of the integral topologies from our analysis. Concretely, the cases $I^{**}_{a,1,1}$ with $a>1$, which correspond to topologies with a bubble correction inserted in one propagator, actually contain a double propagator and thus fall into the integral topology $I^{*}_{a-1,1,1}$, so we do not need to consider these 4 cases. Moreover, $I^{*}_{1,1,1}$ coincides with $I^{**}_{1,1,1}$ due to momentum conservation, leading to a total of 79 integral topologies to be studied; see sec.~\ref{sec: General_analysis_geometries} for the complete list.

\subsection{Kinematics}

In our analysis, we will consider all integrals to depend on generic kinematics, which implies that all internal masses, external masses, and Mandelstam variables are taken to be non-zero and different.\footnote{Additionally, we also require that the kinematic points do not correspond to thresholds (or pseudo-thresholds) of the integrals.} With this setup, the kinematic variables appearing in a given Feynman integral can be separated into two types: internal masses $m_i^2$ and external kinematic variables. Their number depends on the number of propagators $n_{\text{props}}$ and the number of external legs $n_{\text{legs}}$ of the integral, which are given by
\begin{align}
n_{\text{props}} = a+b+c \,, \qquad n_{\text{legs}} = a+b+c-1-n_* \,,
\end{align}
where $n_* \in \{ 0,1,2 \}$ denotes the number of stars in the notation introduced in fig.~\ref{fig:integral_notation}. Concretely, the number of internal masses is equal to $n_{\text{props}}$, whereas the total number of external kinematic variables is set by the number of external masses $p_i^2$ and Mandelstam variables $s_{i_1 \, \cdots \, i_k} \equiv (p_{i_1} + \dots + p_{i_k})^2$, given in tab.\ \ref{eq:kinematicstable}.
\begin{table}[t]
\centering
\begin{tabular}{|l|ccccc|}
\hline
number of legs ($n_{\text{legs}}$) & $0$ & $1$ & $2$ & $3$ & $\geq 4$ \\ \hline
number of external masses & $0$ & $0$ & $1$ & $3$ & $n_{\text{legs}}$ \\
number of Mandelstams & $0$ & $0$ & $0$ & $0$ & $3 n_{\text{legs}} - 10$ \\
\hline
\end{tabular}
\caption{Numbers and types of external kinematic variables for different numbers of external legs.}
 \label{eq:kinematicstable}
\end{table}
Specifically, we see that the number of external kinematic variables equals $4 n_{\text{legs}} - 10$ for $n_{\text{legs}} \geq 4$.

Let us now briefly discuss various specific cases of kinematic configurations. For $n_{\text{legs}}=0$ and $n_{\text{legs}}=1$ there are no external kinematic variables. For $n_{\text{legs}} = 2$, there is only the external mass $p^2$, while for $n_{\text{legs}} = 3$ there are three external masses $\{p_1^2, p_2^2, p_3^2\}$. For $n_{\text{legs}}=4$, there are 4 external masses and 2 independent Mandelstam variables. Yet for explicit computations, particularly for non-planar integrals, it can be convenient to trade one of the external masses and use instead three Mandelstam variables, traditionally defined as
\begin{align}
s \equiv s_{12} \,,\quad t \equiv s_{13} \,,\quad u \equiv s_{14} = p_1^2 + p_2^2 + p_3^2 + p_4^2 - s - t \,. 
\end{align}
For $n_{\text{legs}}=5$, there are 5 external masses and 5 independent Mandelstam variables, which we can choose as adjacent and cyclic, i.e.\
\begin{align}
\{ p_1^2,\; p_2^2,\; p_3^2,\; p_4^2,\; p_5^2,\; s_{12},\; s_{23},\; s_{34},\; s_{45},\; s_{51} \} \,.
\label{eq:fivepointkinematics}
\end{align}

For $n_{\text{legs}} \geq 6$, while the number of variables still follows tab.~\ref{eq:kinematicstable}, there is a complication that did not appear for lower numbers of legs. Concretely, in the 't~Hooft--Veltman scheme we require that external momenta lie in a four-dimensional space, which does not occur automatically for $n_{\text{legs}} \geq 6$. In principle, we can impose that the Gram determinant of a subset of five external momenta is zero, e.g.\ that $\det [G (p_1, p_2, p_3, p_4, p_5)] = 0$, where the Gram matrix is defined via $G_{ij}(\vec{q})=q_i \cdot q_j$. This constraint is included in the counting of tab.~\ref{eq:kinematicstable}, but in practice it can be challenging to impose it in a way that is compatible with our Baikov implementation; cf.~ref.~\cite{Frellesvig:2024ymq}. To this end, we explicitly decompose the external momenta as
\begin{align}
p_i^{\mu} &= a_{i,1} \, p_1^{\mu} + a_{i,2} \, p_2^{\mu} + a_{i,3} \, p_3^{\mu} + a_{i,4} \, p_4^{\mu} \qquad \text{for }\ i=5,\dots,n_{\text{legs}}{-}1 \, ,
\end{align}
which makes the Gram-determinant constraints be satisfied automatically. This way, a natural choice of external kinematic variables are again the 10 variables of 5-point kinematics from eq.~\eqref{eq:fivepointkinematics}, but renamed to
\begin{align}
\{ p_1^2,\; p_2^2,\; p_3^2,\; p_4^2,\; s_{1234},\; s_{12},\; s_{23},\; s_{34},\; s_{123},\; s_{234} \} \, ,
\label{eq:fivepointkinematics2}
\end{align}
along with $4 (n_{\text{legs}} - 5)$ variables $a_{i,j}$. This gives a total number of $4 n_{\text{legs}} - 10$ external kinematic variables, in accordance with the counting of tab.~\ref{eq:kinematicstable}. We stress that the more natural set of $3 n_{\text{legs}} - 10$ Mandelstam variables and $n_{\text{legs}}$ external masses can of course be expressed in terms of the previous variables, but inverting this parametrization is not needed for our purposes.

Let us end this section by discussing the range of kinematic variables in the integrals considered in this paper. The minimum number is $3$, corresponding to the three internal masses of the two-loop tadpole $I_{1,1,1}^{**}$. By contrast, the maximum is $41$ kinematic variables for integrals such as $I_{5,2,4}$, which depend on $11$ internal masses and $30$ external kinematic variables. These $30$ external kinematic variables can be interpreted either as $10$ external masses and $20$ Mandelstam variables, or as the $10$ variables of eq.~\eqref{eq:fivepointkinematics2} along with the $20$ variables $a_{5,j}, \dots, a_{9,j}$ for $j=1,\dots,4$.

\section{Full classification at two loops: General discussion}
\label{sec:classification}

In this section, we review the techniques that can be used to classify Feynman-integral geometries, and perform a general analysis of the geometries that can appear at two-loop order in the 't Hooft--Veltman scheme. First, in sec.~\ref{sec: identifying_geometries}, we review the concepts of Picard--Fuchs operators and leading singularities, focusing on their connection to the underlying Feynman-integral geometries. In sec.~\ref{sec: CICY_geo}, we introduce the construction of manifolds through the complete intersection of hypersurfaces, and show how the associated configuration matrices provide a necessary but not sufficient criteria for the occurrence of non-trivial geometries at the level of the maximal cut. Then, in sec.~\ref{sec: LBL_Baikov}, we briefly introduce the loop-by-loop Baikov representation for Feynman integrals and emphasize its usefulness for computing leading singularities. Lastly, in sec.~\ref{sec: General_analysis_geometries}, we bound the geometries that can appear at two-loop order, based uniquely on the result of the maximal cut of the integrals. We will look at all integral topologies where this upper bound is not saturated or where further subtleties occur in sec.\ \ref{sec:special_cases}.

\subsection{Identifying geometries in Feynman integrals}
\label{sec: identifying_geometries}

All Feynman integrals can be expressed in terms of integrated integrals. One important feature that distinguishes them, however, is the geometry over which these iterated integrals are defined.\footnote{Strictly speaking, in the context of Feynman integrals one encounters varieties, not manifolds, since the associated geometries are typically singular. However, we will not make such a distinction in the following, since it is not necessary to desingularize the geometries for the purpose of evaluating the Feynman integrals.} Apart from the Riemann sphere, Feynman integrals have been observed to involve elliptic curves, higher-genus hyperelliptic curves as well as CY geometries; see ref.~\cite{Bourjaily:2022bwx} for a recent review. In general, there exist two methods that allow us to detect such geometries before carrying out the full evaluation of the integrals:
\begin{itemize}
\item Investigating the Picard--Fuchs operator of the integrals;
\item Analyzing the leading singularity of the integrals.
\end{itemize}
Let us now briefly discuss these two methods. We refer to refs.~\cite{Frellesvig:2024zph, Brammer:2025rqo} for further details, where a subset of the authors used similar methods to investigate the geometries appearing in classical gravity.

A \emph{Picard--Fuchs operator} is a differential operator with respect to a kinematic variable that is associated to a given Feynman integral, with the property that applying the operator to the integral only yields its subsectors, i.e.\ integrals with a smaller number of propagators. Typically, a Picard--Fuchs operator factorizes into a product of rational lower-order operators, with each operator within the factorization characterizing an aspect of the associated Feynman-integral geometry~\cite{Adams:2017tga}. For example, if the Picard--Fuchs factorizes completely into first-order operators, and the same occurs iteratively for all subsectors, the integral is guaranteed to admit a $d$log form. Feynman integrals that admit a $d$log form can typically be expressed in terms of multiple polylogarithms; see however ref.\ \cite{Duhr:2020gdd} for a (non-Feynman integral) counterexample. 
Instead, if in addition the factorization of the Picard--Fuchs operator contains an irreducible second-order operator, the integral can at most be elliptic, and similarly for higher orders. Then, by studying the properties of these irreducible operators, one can characterize the precise geometry at hand; see e.g.\ refs.~\cite{Pogel:2022vat,Frellesvig:2024rea} for a discussion on CY operators. Let us note that, for multi-scale Feynman integrals, the factorization of Picard--Fuchs operators becomes more intricate, as square roots in the kinematic variables can appear. Thus, even for integrals admitting a $d$log form, a rational factorization may not be possible; see app.~\ref{app:factorization_PF} for an example. 

While studying the Picard--Fuchs operator of a Feynman integral suffices to prove the presence of a non-trivial geometry, obtaining this operator can be computationally very expensive. In practice, it commonly involves solving the integration-by-parts (IBP) identities that Feynman integrals satisfy~\cite{Chetyrkin:1981qh,Tkachov:1981wb}, which allows to express any Feynman integral in a sector in terms of a minimal subset of independent integrals, the so-called master integrals. Even though highly optimized computer implementations for solving IBP relations exist, such as those in FIRE~\cite{Smirnov:2025prc} and Kira~\cite{Lange:2025fba},\footnote{See also ref.~\cite{Smirnov:2025dfy} for a recent review.} solving them becomes increasingly challenging at high loop orders and especially in cases with multiple kinematic variables, such as the integrals considered in this paper.\footnote{Still, let us highlight refs.~\cite{vonHippel:2025okr,Song:2025pwy,Zeng:2025xbh} for recent improvements on IBP reduction assisted by machine learning, as well as refs.~\cite{Muller-Stach:2012tgj,Lairez:2022zkj,delaCruz:2024xit} for
an alternative method to obtain Picard--Fuchs operators that does not rely on IBPs.} Consequently, we will use a different and computationally much lighter approach based on the so-called maximal cut to characterize the underlying geometries, which we turn to next.

A \emph{generalized cut}~\cite{Britto:2024mna} is a deformation of the integration contour of a Feynman integral such that it encircles the point where a given propagator vanishes, effectively computing the on-shell residue. In practice, generalized cuts are most conveniently performed in the so-called Baikov representation, which we will introduce in sec.~\ref{sec: LBL_Baikov}. The \emph{maximal cut} consists of performing such generalized cuts for all propagators of the integral. Then, the so-called \emph{leading singularity}~\cite{Cachazo:2008vp,Arkani-Hamed:2010pyv} corresponds to deforming all remaining integration contours to closed contours. In cases where these deformed contours encircle poles, this corresponds to taking all possible further residues of the maximal cut, which for practical purposes is conveniently implemented in the {\DlogBasis} package~\cite{Henn:2020lye} in {\texttt{\textup{Mathematica}}}. 

Since cutting propagators commutes with taking derivatives with respect to the kinematic variables, the leading singularity is annihilated by the Picard--Fuchs operator of the integral, and thus characterizes the Feynman-integral geometry too~\cite{Primo:2016ebd,Bosma:2017ens}. Consequently, if all leading singularities of a Feynman integral are algebraic (i.e.\ there are no integrals remaining after taking residues), the corresponding Picard--Fuchs operator is a product of first-order operators, and the Feynman integral is guaranteed to admit a $d$log form on the maximal cut. Otherwise, the integrals expressing the leading singularity are periods of the underlying geometry, and the task is then to find changes of variables that make the nature of this geometry manifest; see sec.~\ref{sec: LBL_Baikov} for details. In many cases, the result of the leading singularity is an $n$-fold integral such as
\begin{align}
\int \frac{dz_1 \cdots dz_n}{\sqrt{P_m(z_1,\ldots,z_n)}}\,,
\label{eq:geodef}
\end{align}
where $P_m$ is a polynomial of degree $m$. The distinguishing feature between the various geometries is then the number $n$ of transcendental integrals remaining, and the polynomial $y^2 = P_m(z_1,\ldots,z_n)$ appearing in the denominator. In tab.~\ref{tab:geometries}, we summarize the characteristic leading singularity for different non-trivial geometries relevant to this work.\footnote{In general, these geometries can also be defined through odd polynomials with one degree less, e.g.\ an elliptic curve is also given by $y^2=P_3(z)$.} Let us note, however, that the expression in eq.~\eqref{eq:geodef} is not the only way that a non-trivial geometry can arise, as it assumes that we have a single polynomial and one square root. As we will discuss in sec.~\ref{sec: CICY_geo}, one can also obtain non-trivial geometries through the complete intersection of hypersurfaces defined by various polynomials.

\begin{table}
\begin{center}
\begin{tabular}{|l|l|}
\hline
Geometry & Characteristic equation \\ \hline
Elliptic curve & $y^2=P_4(z)$ \\
Hyperelliptic curve (genus $g$) & $y^2=P_{2g+2}(z)$ \\
Del Pezzo surface of degree 2 & $y^2=P_4(z_1,z_2)$ \\
K$3$ surface & $y^2=P_6(z_1,z_2)$ \\
Calabi--Yau $n$-fold & $y^2=P_{2n+2}(z_1,\ldots,z_n)$ \\
\hline
\end{tabular}
\end{center}
\caption{Summary of the various Feynman-integral geometries that we will encounter in this work, along with their characteristic polynomial from eq.~\eqref{eq:geodef}.}
\label{tab:geometries}
\end{table}

While there is essentially a unique contour encircling the pole where a given propagator vanishes, there is typically more than one possible closed contour for the remaining integration variables after having taken the maximal cut. Thus, there is not a unique leading singularity. Consequently, we need to analyze all leading singularities arising from different closed contours, as described by homology. 
A prototypical example of this is an elliptic integral of the third kind,
\begin{align}
\label{eq:elliptic_third_kind}
\int \frac{d z}{(z - a) \sqrt{(1-z^2) (1 - k^2 z^2)}}\,.
\end{align}
There is a closed contour encircling the pole at $z=a$, with a corresponding algebraic leading singularity. However, there are also two independent closed contours that pass, encircle or go through branch cuts of the square root, corresponding to the A and B cycle of the torus associated to the elliptic curve. The leading singularities corresponding to the latter contours are transcendental integrals that reveal the presence of the elliptic curve.

Alternatively to homology, we can look at cohomology, i.e.~at the basis of integrands (which define the different master integrals). Thus, in addition to the form in eq.\ \eqref{eq:elliptic_third_kind}, we can also have an integrand with an additional numerator $(z-a)$, which cancels the corresponding denominator such that the pole at $z=a$ is absent. In such a case, we only have leading singularities given by elliptic integrals. Since homology is more challenging to visualize and interpret in the case of higher dimensions, in this paper we will usually take the point of view of cohomology, i.e.\ consider all possible (master) integrals in a topology or sector.

\subsection{Complete intersection manifolds}
\label{sec: CICY_geo}

Quite often, computing the maximal cut does not directly yield a form such as eq.~\eqref{eq:geodef}, where there is a single polynomial under the square root. Instead, the result commonly involves a denominator with several polynomials and different square roots, which requires introducing non-trivial changes of variables and rationalizations to attain eq.~\eqref{eq:geodef}; see sec.~\ref{sec: LBL_Baikov} for details. Notably, finding such changes of variables becomes a particularly challenging task. Thus, to reduce the complexity of our analysis, we aim to detect the non-trivial geometries at the level of the maximal cut -- with possibly multiple polynomials and square roots. Still, in practice we explicitly calculate the leading singularities using changes of variables and rationalizations for all integrals, but knowing beforehand which cases may actually involve non-trivial geometries simplifies this task considerably.

With this goal in mind, let us briefly introduce the notion of complete intersection (Calabi--Yau) manifolds~\cite{Green:1986ck,Candelas:1987kf}; see ref.~\cite{Hubsch:1992nu} for a pedagogical introduction to the topic. In general, the intersection of $k$ different hypersurfaces $X^a$ in an embedding space $\mathcal{X}$ defines a manifold $\mathcal{M}$,
\begin{equation}
\mathcal{M} = \cap_{a=1}^k X^a \subset \mathcal{X}\,.
\end{equation}
In particular, we have a complete intersection manifold if the hypersurfaces meet transversely, such that there are no points of degeneracy where two $X^a$ simply touch. Then, the dimension of a complete intersection manifold is given by $\dim \mathcal{M} = \dim \mathcal{X} - k$. For example, the complete intersection of two spheres produces a circle.

For the case of interest, let us consider an embedding space $\mathcal{X}$ given by $n$-dimensional complex projective space $[z_1: \dots : z_{n+1}] \sim [\lambda z_1: \dots : \lambda z_{n+1}] \in \mathbb{P}^n$ for $\lambda\in \mathbb{C}\backslash \{ 0 \}$. Then, we can define the hypersurfaces $X^a$ as the zero loci of homogeneous polynomials $f_a$ of degree $q_a$ with respect to different variables $z_{a_1},\dots,z_{a_r}$ in $\mathbb{P}^n$,\footnote{In general, the embedding space can decompose into a product of spaces -- commonly a product of projective spaces $\mathbb{P}^{n_1} \times \dots \times \mathbb{P}^{n_m}$ -- since different sets of polynomials can depend on a disjoint subset of variables $z_i$; see refs.~\cite{Green:1986ck,Candelas:1987kf,Hubsch:1992nu} for details. In our case, however, the polynomials generically depend on all variables $z_i$, and no decomposition into smaller subspaces occurs.}
\begin{equation}
X^a: \quad f_a(z_{a_1},\dots,z_{a_r}) = 0\,, \quad \text{ for } \ a=1,\dots,k\,.
\end{equation}
Now, we can define the $1 \times k$ configuration matrix\footnote{In the case of an embedding space given by the product of $m$ projective spaces, we actually have an $m \times k$ matrix, not just one single row.}
\begin{equation}
\mathcal{M} \in [\mathbb{P}^n \, || \, q_1 \ \cdots \ q_k ]\,,
\end{equation}
which specifies the embedding space and the degrees of homogeneity of the polynomials. Then, we have a (complete intersection) Calabi--Yau $(n-k)$-fold if~\cite{Green:1986ck,Candelas:1987kf,Hubsch:1992nu}
\begin{equation}
\label{CY_condition_P}
n+1 = \sum_{a=1}^k q_a\,,
\end{equation}
which can be easily checked at the level of the configuration matrix. For example, we can have the configuration matrices
\begin{equation}
[\mathbb{P}^4 \, || \, 5], \qquad [\mathbb{P}^5 \, || \, 3 \ 3], \qquad [\mathbb{P}^5 \, || \, 4 \ 2], \qquad [\mathbb{P}^6 \, || \, 3 \ 2 \ 2], \qquad [\mathbb{P}^7 \, || \, 2 \ 2 \ 2 \ 2].
\end{equation}
All of these cases satisfy the CY condition~\eqref{CY_condition_P} and have $\dim \mathcal{M}=3$; thus, they correspond to CY threefolds.

In the case of Feynman-integral geometries defined through the maximal cut, however, we also have square roots. To account for them, we need to generalize the embedding space to weighted projective space; see ref.~\cite{Bourjaily:2019hmc} for a discussion in the context of Feynman integrals. In particular, we now allow for different weights $w_i$ for each coordinate $z_i$,
\begin{equation}
[z_1: \dots : z_{n+1}] \sim [\lambda^{w_1} z_1: \dots : \lambda^{w_{n+1}} z_{n+1}] \in \mathbb{WP}^{w_1,\dots,w_{n+1}}\,.
\end{equation}
For example, for a homogeneous degree-8 polynomial, the equation $y^2 = P_8(z_1,\dots,z_4)$ would be defined in weighted projective space $[z_1:z_2:z_3:z_4:y] \in \mathbb{WP}^{1,1,1,1,4}$, where the weight of $y^2$ is adjusted to match the total weight of the polynomial. Then, we have a CY $(n-k)$-fold if the following condition holds:
\begin{equation}
\label{CY_condition_WP}
\sum_{i=1}^{n+1} w_i = \sum_{a=1}^k q_a \,.
\end{equation}
Notice that when all $w_i=1$, this condition reduces to eq.~\eqref{CY_condition_P}, since $\mathbb{WP}^{1,1,\dots,1} = \mathbb{P}^{n}$. For example, the previous case with a homogeneous degree-8 polynomial defines a configuration matrix
\begin{equation}
[\mathbb{WP}^{1,1,1,1,4} \, || \, 8]\,,
\end{equation}
which clearly satisfies the CY condition for a CY threefold. Similarly, we can consider the defining equations for an elliptic curve, K3 surface and CY $n$-fold presented in tab.~\ref{tab:geometries}. In these cases, we must first homogenize the polynomials by introducing one extra variable $z_0$, such as $\widetilde{P}_4(z_0,z)=z_0^4 \, P_4(z/z_0)$; then, we have the configuration matrices
\begin{equation}
[\mathbb{WP}^{1,1,2} \, || \, 4]\,, \qquad [\mathbb{WP}^{1,1,1,3} \, || \, 6]\,, \qquad [\mathbb{WP}^{1,\dots,1,1,n+1} \, || \, 2n+2]\,.
\end{equation}
One can check that they satisfy the CY condition~\eqref{CY_condition_WP} and define a CY of the correct dimension, respectively.

Calabi--Yau geometries lie on the line between trivial and non-trivial geometries. If the combined degrees of the polynomials defining the geometry are lower than for a Calabi--Yau, we have Fano varieties, which are known to be rationalizable. Calabi--Yau geometries are the first non-rationalizable geometries one encounters with increasing degree. Geometries defined by polynomials with higher combined degree are non-rationalizable and more complicated than Calabi--Yaus. In general, this is part of the so-called Enriques--Kodaira classification of compact complex surfaces, which is based on their Kodaira dimension; see e.g.~ref.~\cite{Barth2004} for an introduction. In the context of Feynman integrals, all known examples depending on non-rationalizable geometries either involve a Calabi--Yau or a hyperelliptic curve, which is a particular instance of a one-dimensional geometry of general type. At the time of writing, no higher-dimensional geometries of general type have been identified in the context of Feynman integrals.

Now, the idea is to use the framework of complete intersection manifolds to detect potential non-trivial geometries directly from the maximal cut. Concretely, we can collect all combinations of hypersurfaces defined by the various polynomials and square roots in the denominator of the maximal cut, assemble the configuration matrix, and check whether the degrees satisfy or exceed the CY condition~\eqref{CY_condition_WP}. Whenever this is the case, the integral is a candidate for having a non-trivial geometry, which provides us with a guiding principle for finding the appropriate changes of variables. By contrast, if the sum of degrees is too small to meet the CY condition, it indicates that the corresponding geometry is rationalizable.
  
At this point, it is important to stress the limitations and scope of this approach. The configuration matrix is insensitive to the details -- such as singularities or degeneracies -- of the polynomials involved, which can reduce the complexity of the geometry. As such, the configuration matrix only provides a necessary but not a sufficient criterium for the occurrence of non-trivial geometries. We will frequently encounter cases where an elliptic curve indicated by the configuration matrix is not actually present since the polynomial under the square root is a perfect square; see e.g.\ secs. \ref{subsubsec:5Tbox} and \ref{subsubsec:doublebox}. 

Our general procedure is to first check for non-trivial geometries of the highest possible dimension. If they are absent, we proceed to check for non-trivial geometries of dimension one lower, and so on, up to dimension 1. Including polynomials not associated to square roots corresponds to taking residues at the poles of these polynomials, which is only indicative of the geometry in the absence of non-trivial geometries in the other factors; cf.\ the discussion at the end of sec.~\ref{sec: identifying_geometries}.

To exemplify the application of configuration matrices, let us consider two concrete examples, where the maximal cut is derived through the loop-by-loop Baikov representation; see sec.~\ref{sec: LBL_Baikov} for details. First, let us study the corner integral for the integral topology $I_{1,1,1}$ in $D=2$, i.e.\ the integral where all propagators occur with unit powers and the numerator is $1$. This corresponds to the well-studied elliptic sunrise integral~\cite{Sabry:1962rge,Broadhurst:1993mw,Laporta:2004rb,Adams:2013nia,Bloch:2013tra,Adams:2014vja,Broedel:2017siw,Giroux:2022wav}. Concretely, on the maximal cut, we have
\begin{equation}
\label{eq:before_config_matrix_sunrise}
\twoloop[0]{1}{1}{1} \, \Bigg|_{\text{max-cut}} \propto \int \frac{dz}{\sqrt{P_2(z)} \sqrt{Q_2(z)}} \, , 
\end{equation}
where $P_2$ and $Q_2$ are quadratic polynomials. Introducing the homogenization variable $z_0$, and denoting the square roots as $y_1$ and $y_2$, respectively, we have the configuration matrix
\begin{equation}
\label{eq:config_matrix_sunrise}
[\mathbb{WP}^{1,1,1,1} \, || \, 2 \ 2]
\end{equation}
in weighted projective space $[z_0:z:y_1:y_2] \in \mathbb{WP}^{1,1,1,1}$. As expected, the sunrise integral satisfies the CY condition~\eqref{CY_condition_WP} for an elliptic curve. In particular, by rationalizing either of the square roots using eq.~\eqref{eq: change_of_variables_(z-r1)(z-r2)}, one can rewrite the result of the maximal cut as in eq.~\eqref{eq:geodef}, which manifests the ellipticity.\footnote{Naively combining the two square roots, on the other hand,  yields a different but isogenous elliptic curve \cite{Frellesvig:2021vdl}.} 
Let us note, however, that focusing solely on the corner integral is in general not sufficient to characterize the Feynman-integral geometry in a given sector. In particular, there is an additional polynomial, which has vanishing exponent in $D=2$ for the corner integral of the sunrise. Importantly, this polynomial can appear for other master integrals, and thus induce a different geometry; see sec.~\ref{sec: LBL_Baikov} for a discussion. However, in the case of the sunrise, this additional polynomial only adds a marked point on the elliptic curve, and acts analogously as in the example from eq.~\eqref{eq:elliptic_third_kind}.

Similarly, we can consider the corner integral for the integral topology $I_{2,2,2}$ in $D=4$, which corresponds to the 5-point tardigrade integral~\cite{Bourjaily:2018yfy}. On the maximal cut, we obtain
\begin{equation}
\label{eq:tardigrade_example}
\twoloop[0]{2}{2}{2} \, \Bigg|_{\text{max-cut}} \propto \int \frac{dz_1 dz_2 dz_3}{P_2(z_1,z_2,z_3) \sqrt{Q_4(z_1,z_2,z_3)}} \, , 
\end{equation}
where $P_2$ and $Q_4$ are quadratic and quartic polynomials, respectively. Focusing first on the square root $y^2=Q_4(z_1,z_2,z_3)$, we can introduce the homogenization variable $z_0$, and obtain the configuration matrix
\begin{equation}
\label{eq:tardigrade_example_config_matr0}
[\mathbb{WP}^{1,1,1,1,2} \, || \, 4]
\end{equation}
in weighted projective space $[z_0:z_1:z_2:z_3:y] \in \mathbb{WP}^{1,1,1,1,2}$. Since the degree is smaller than the sum of projective weights, it shows that $\sqrt{Q_4}$ alone does not introduce a non-trivial geometry.
We thus proceed to include $P_2$, yielding 
\begin{equation}
\label{eq:tardigrade_example_config_matr}
[\mathbb{WP}^{1,1,1,1,2} \, || \, 2 \ 4] \,.
\end{equation}
Now, this case satisfies the CY condition~\eqref{CY_condition_WP} for a K3 surface, as already shown using a different approach in refs.~\cite{Lairez:2022zkj,Doran:2023yzu}. Once again, via changes of variables one can rewrite the leading singularity as in eq.~\eqref{eq:geodef}, which makes the K3 surface explicit; we refer to sec.~\ref{subsubsec:tardigrade} for details. As in the previous example, to entirely characterize the geometry in this sector, one should also consider the additional polynomials appearing in other master integrals; cf.~sec.~\ref{subsubsec:tardigrade}.

Further examples can be found in sec.~\ref{sec:special_cases}, where we explicitly provide the configuration matrices for a selection of Feynman integrals of interest.

\subsection{Loop-by-loop Baikov representation at two loops}
\label{sec: LBL_Baikov}

The computation of the maximal cut and the leading singularity can be streamlined using the so-called Baikov representation of Feynman integrals~\cite{Baikov:1996iu}. The defining feature of the Baikov representation is that the propagators of the integral in momentum representation become the integration variables, the so-called Baikov variables. Thus, generalized cuts can be performed using a simple residue operation in the Baikov variables, making this representation well suited for our current analysis. In this section, we briefly review the Baikov representation at two loops; see ref.~\cite{Frellesvig:2024ymq} for an introduction.

While the Baikov representation is generally effective for computing the maximal cut of an integral, one downside is that it requires additional integration variables apart from the propagators in many cases beyond one-loop order. These extra variables correspond to Irreducible Scalar Products (ISPs), and they are required to promote the set of propagators to integration variables. The ISPs are not affected by the maximal cut; thus, it is desirable to find a Baikov representation that minimizes the number of ISPs, as it will naturally reduce the complexity of the integral representation. This can be done using the so-called loop-by-loop version of the Baikov parametrization~\cite{Frellesvig:2017aai}, and a particular loop-by-loop ordering minimizing the number of ISPs can be chosen~\cite{Frellesvig:2024ymq}. 

For two-loop Feynman integrals with a fixed parametrization of the loop momenta, there exist two options for loop-by-loop orderings. 
As discussed in ref.~\cite{Frellesvig:2024ymq}, the option that yields the lowest number of ISPs is the one starting with the loop with fewest propagators. In our case, it always corresponds to the loop in fig.~\ref{fig:integral_notation} with $b+c$ propagators, and we denote its loop momentum as $k_1$. Then, under the loop-ordering $\{ k_1, k_2\}$, we have a total of
\begin{align}
n_B = 2 + E_1 + E_2
\label{eq:nvars}
\end{align}
Baikov variables, which we will collectively denote as $\vec{z}$. Here,
\begin{equation}
E_1 = \min\{(b-1)+(c-1)+1 , \, 5\} = \min\{b+c-1 , \, 5\}
\end{equation}
and 
\begin{align}
E_2 &= \min\{ \max\{(a-1)+(b-1)+(c-1)+(2-n_*)-1,0\} , \, 4\} \nonumber \\
&= \min \{ \max\{a+b+c-2-n_*,0\} , \, 4 \}
\end{align}
denote the number of independent external momenta relative to the $k_1$-loop and to the entire integral, respectively, where we recall that $n_* \in \{ 0,1,2 \}$ is the number of stars in the integral.%
\footnote{Note that the first loop can have up to five linearly independent external momenta since the second loop momentum is external to it and is not constrained to live in four dimensions.} Given these definitions, the number of ISPs is obtained by subtracting the number of propagators from the number of Baikov variables,
\begin{equation}
\nISP = n_B-a-b-c\, .
\end{equation}

With this, the loop-by-loop Baikov representation at two-loop order becomes~\cite{Frellesvig:2024ymq}
\begin{align}
I \, = \, - \frac{\mathcal{J}}{\pi^{(n_B - 1)/2}} \frac{\mathcal{E}_2^{\gamma_2}}{\Gamma \big( \tfrac{D{-}E_{1\!}}{2} \big) \Gamma \big( \tfrac{D{-}E_{2\!}}{2} \big) } \! \int_{\mathcal{C}} \frac{d^{n_B} z \ \mathcal{N}(\vec{z})}{z_1 \cdots z_{a+b+c}} \, \mathcal{B}_2(\vec{z})^{\beta_2} \, \mathcal{E}_1(\vec{z})^{\gamma_1} \, \mathcal{B}_1(\vec{z})^{\beta_1} \, ,
\label{eq:Baikov}
\end{align}
where we have suppressed potential propagator powers and included a numerator factor $\mathcal{N}(\vec{z})$ that can depend on all Baikov variables. Moreover, $\mathcal{J}$ denotes a constant Jacobian, which is given by $\mathcal{J} = \pm 2^{2-n_B}$ depending on the exact expressions for the propagators. Introducing $G$ as the Gram matrix, defined as $G_{ij}(\vec{q})= q_i \cdot q_j$, the so-called Baikov polynomials are given by
\begin{align}
\label{eq:Bdef_1}
\mathcal{E}_2 & \equiv \det G(\bar{p}_2) \,, & \mathcal{B}_2 & \equiv \det G(k_2, \bar{p}_2) \,, \\
\mathcal{E}_1 & \equiv \det G(k_2, \bar{p}_1) \,,& \mathcal{B}_1 & \equiv \det G(k_1, k_2, \bar{p}_1) \,,
\label{eq:Bdef_2}
\end{align}
while their zero loci provide the boundary for the integration domain (or chamber) $\mathcal{C}$.
Here, we use $\bar{p}_1$ to denote the $(E_1-1)$ independent external momenta $p_i$ appearing in the first loop, and $\bar{p}_2=\{ p_1, \dots, p_{E_2}\}$ is the set of $E_2$ independent external momenta of the entire integral. The exponents of the Baikov polynomials in eq.~\eqref{eq:Baikov} are given by
\begin{equation}
\label{eq:gammadef}
\gamma_i \equiv \frac{E_i-D+1}{2} \, , \qquad \qquad \beta_i \equiv \frac{D-E_i-2}{2} \, .
\end{equation}

Furthermore, the Gram determinants appearing in the Baikov polynomials satisfy the Desnanot--Jacobi identity~\cite{Dlapa:2021qsl, Chen:2022lzr, Frellesvig:2024ymq}. Denoting by $M$ the determinant of a Gram matrix, and by $M^{i_1,\dots,i_n}_{j_1,\dots,j_m}$ the determinant of the matrix where we remove the rows $i_1,\dots,i_n$ and columns $j_1,\dots,j_m$, the Desnanot--Jacobi identity reads
\begin{equation}
\label{eq: Desnanot_Jacobi}
M^n_n \, M^1_1 = (M^1_n)^2 + M \, M^{1,n}_{1,n}\,.
\end{equation}
Therefore, from the definitions of the Baikov polynomials in eqs.~\eqref{eq:Bdef_1}--\eqref{eq:Bdef_2}, and taking $M=\mathcal{B}_1 \equiv \det G(k_1, \bar{p}_1, k_2)$, we have the relation
\begin{equation}
\det G(k_1,\bar{p}_1) \, \mathcal{E}_1 = (M^1_n)^2 + \det G(\bar{p}_1) \, \mathcal{B}_1\,,
\end{equation}
where $\mathcal{E}_1 \equiv \det G(\bar{p}_1, k_2)$ is another Baikov polynomial, and $M^1_n$ is the corresponding determinant where we remove the first row and the last column. Then, if $\mathcal{E}_1$ or $\mathcal{B}_1$ vanish, they respectively imply
\begin{equation}
\label{eq: degeneration_Desnanot_Jacobi}
\mathcal{E}_1=0 \enspace \Longrightarrow \enspace \mathcal{B}_1 = \frac{- (M^1_n)^2}{\det G(\bar{p}_1)}\,, \qquad \qquad \mathcal{B}_1=0 \enspace \Longrightarrow \enspace \mathcal{E}_1 = \frac{(M^1_n)^2}{\det G(k_1,\bar{p}_1)}\,.
\end{equation}
Importantly, for planar topologies, both $\det G(\bar{p}_1)$ and $\det G(k_1,\bar{p}_1)$ are constant after the maximal cut since they do not depend on ISPs. As a consequence, for planar topologies, the zero locus of $\mathcal{E}_1$ or $\mathcal{B}_1$ enforces that the other polynomial becomes a function of the ISPs that is a perfect square. We will use this property several times throughout this paper, starting in sec.~\ref{subsubsec:5Tbox}. Note that for non-planar topologies, at least one of the scalar products $k_1\cdot p_i$ will depend on the ISPs, which spoils the perfect square in $\mathcal{E}_1$ in the right part of eq.~\eqref{eq: degeneration_Desnanot_Jacobi}; see sec.~\ref{subsubsec:I323} for an example.

The main advantage of the Baikov representation is that performing the maximal cut becomes trivial: we simply take the residue at the poles where the Baikov variables that correspond to propagators vanish, i.e.\ for $\{ z_1, \dots, z_{a+b+c} \}=0$. The result for the case of unit propagator powers is then
\begin{align}
I_{\text{max-cut}} \, \propto \int_{\mathcal{C}} d^{\nISP} z \ \mathcal{N}(\vec{z}) \, \mathcal{B}_2(\vec{z})^{\beta_2} \, \mathcal{E}_1(\vec{z})^{\gamma_1} \, \mathcal{B}_1(\vec{z})^{\beta_1} \, ,
\label{eq:max_cut_Baikov}
\end{align}
where the Baikov polynomials are implicitly understood to be evaluated at $z_1=\dots=z_{a+b+c}=0$, and where we have dropped the constant prefactors. As can be seen, after the maximal cut, $\nISP$ integrations remain over the ISPs $\{ z_{a+b+c+1}, \dots, z_{n_B} \}$, which bounds the dimension of the underlying geometry. 
Note that the Baikov polynomials $\mathcal{B}_1$, $\mathcal{B}_2$ and $\mathcal{E}_1$ correspond to the (non-homogenized) polynomials in the configuration-matrix approach, and that additional variables $y_i$ are introduced in the case of square roots.

In practice, the limit $d \rightarrow 4$ often reveals further simple poles in eq.~\eqref{eq:max_cut_Baikov}, and if the integral does not contain a non-trivial geometry in their absence, the leading singularity thus corresponds to taking the residue at those poles too, which reduces the complexity of the geometry. 

Such simple poles may only be apparent, however, after performing changes of variables, e.g.\ through the rationalization of square roots. In particular, for most of the analysis at two loops, we find it sufficient to use $\text{GL}(n,\mathbb{C})$ rotations as well as two particular variable transformations already employed in the analysis of Feynman-integral geometries relevant to black-hole scattering up to four loops~\cite{Frellesvig:2024zph,Brammer:2025rqo}. First of all, for square roots of quadratic polynomials such as $\sqrt{(z_i-r_1)(z_i-r_2)}\,$, where $r_j$ are the roots, one can use the variable transformation from $z_i$ to $x_i$~\cite{Bonciani:2010ms,Adams:2018yfj}
\begin{equation}
\label{eq: change_of_variables_(z-r1)(z-r2)}
z_i = r_1-\frac{(r_2-r_1)(1-x_i)^2}{4x_i}\,.
\end{equation}
Secondly, for square roots such as $\sqrt{z_i-r^2}$, which have a perfect square as a root, one can use the rationalization
\begin{equation}
\label{eq: change_of_variables_(z-c_squared)}
z_i = \frac{1-2i r \, x_i}{x_i^2}\,,
\end{equation}
which depends only linearly on $r$. Note that we could have chosen to use $+2irx_i$ instead of $-2irx_i$ in order to rationalize $\sqrt{z_i-r^2}$. A priori, this choice seems to break the Galois symmetry $i\rightarrow-i$ that any physical quantity should be invariant under, and so one could expect the resulting leading singularity to characterize the geometry only up to a Galois orbit. However, in all examples for which we have used this change of variables, the resulting polynomials (which have explicit coefficient dependence on $i$) become Galois invariant via the simple transformation $x_i\rightarrow i\, x_i$, manifesting that the associated geometry is unique; see sec.~\ref{subsubsec:5Tbox} for an example. Besides these variable transformations, we only require the rationalization of a Del Pezzo surface of degree 2, see sec.~\ref{sec:special_case_goomba} as well as app.~\ref{app:rationalization_Del_Pezzo} for details. 

Notably, the transformation from eq.~\eqref{eq: change_of_variables_(z-r1)(z-r2)} actually allows us to rationalize square roots over certain polynomials of high degree when multiple integration variables are involved. Concretely, let us consider an $n$-fold integral over a single square root $y^2=P_m(z_1,\dots,z_n)$, such as in eq.~\eqref{eq:geodef}. Then, if $P_m$ is a polynomial of overall degree $m\leq 2n$ and it is at most quadratic in all variables, we actually have only algebraic leading singularities. Let us now show that this is the case. First, changing variables from $z_1$ to $x_1$ using eq.~\eqref{eq: change_of_variables_(z-r1)(z-r2)}, we simply obtain
\begin{equation}
\label{eq:rationalization_quadratic_polynomials}
- \int \frac{dx_1 dz_2 \dots dz_n}{x_1 \, \sqrt{a_{1,2}(z_2,\dots,z_n)}}\,,
\end{equation}
where $a_{1,2}$ is the coefficient of $z_1^2$ in $P_m$. Importantly, $a_{1,2}$ is itself a polynomial of degree $\leq 2(n-1)$, and at most quadratic in all $n-1$ variables. Therefore, we can perform again the change of variables~\eqref{eq: change_of_variables_(z-r1)(z-r2)} to rationalize its square root, in this case with respect to $z_2$. This way, we can change variables recursively, rationalize all square roots and ultimately obtain an algebraic leading singularity.

Although the changes of variables above allow us to expose simple poles, as explained in sec.~\ref{sec: identifying_geometries}, we must be careful when taking residues in the presence of non-trivial geometries. This is because the integral topology, or sector, also contains (master) integrals where those simple poles are canceled by the numerator factor $\mathcal{N}(\vec{z})$; recall the example from eq.~\eqref{eq:elliptic_third_kind}. To ensure that the maximal cut does not contain a non-trivial geometry -- and that taking a residue is thus legitimate -- we rely on the analysis of complete intersection manifolds from sec.~\ref{sec: CICY_geo}. Concretely, we compute the configuration matrix obtained from the maximal cut of the corner integral, and determine whether a non-trivial geometry is involved. 

A further subtlety arises because some Baikov polynomials in eq.~\eqref{eq:gammadef} have vanishing exponents for the corner integral in four dimensions. Nevertheless, they can reappear with integer exponent at higher orders in the $\varepsilon$-expansion and for other master integrals, for instance when a propagator is dotted, and introduce a different geometry in cases with 2 or more ISPs. An example where this happens is the integral topology $I_{2,1,2}$, also known as the 4-point kite, which is analyzed in detail in sec.~\ref{subsubsec:kite}. To capture any geometry that may arise from these other polynomials, we compute the configuration matrices for all possible combinations of Baikov polynomials, including those with vanishing exponents. Then, only in the cases for which all configuration matrices detect no geometries, and all square roots can be simultaneously rationalized, can residues at simple poles be taken safely. In fact, different polynomial combinations may lead to different geometries. 
However, in all cases we encounter in this paper, we find a unique non-trivial geometry of highest dimension. If we show that such a most intricate geometry is present (e.g.\ through changes of variables), then simple poles associated with any of the involved polynomials cannot be used for taking residues. Similarly, we cannot take poles at simple poles of other polynomials, since we would have the analog of an elliptic integral of the third kind; recall eq.~\eqref{eq:elliptic_third_kind}. 
These simple poles thus lead to marked points on the non-trivial geometry of maximal dimension if its dimension is one, to marked points and marked curves on the non-trivial geometry of maximal dimension if its dimension is two, and so on.  
With this approach, we characterize the geometries for any integral pertaining to a given topology or sector, including (master) integrals with any number of dots and ISPs in the numerator.

\subsection{General analysis of geometries at two loops}
\label{sec: General_analysis_geometries}

Having introduced the Baikov representation, let us now present a general analysis of the geometries that can appear at two-loop order. Specifically, we focus on the result of the maximal cut for each of the 79 integral topologies from sec.~\ref{sec:intro_basis_integrals}. 
This analysis leads to an upper bound on the involved geometry that is satisfied in many cases. We treat cases where it is not satisfied or where further subtleties occur in sec.\ \ref{sec:special_cases}. 
Importantly, this analysis only depends on the numerator $\mathcal{N}(\vec{z})$ insofar as it can cancel poles in the denominator. Our analysis thus holds for all integrals in the topology or sector, and in particular for all master integrals.

As a starting point, we note that the exponents of the three Baikov polynomials $\mathcal{B}_2$, $\mathcal{E}_1$, $\mathcal{B}_1$ in the Baikov representation are given by eq.~\eqref{eq:gammadef} as
\begin{align}
(D-E_2-2)/2 \,,\quad\; (E_1-D+1)/2 \,,\quad\; (D-E_1-2)/2\,,
\end{align}
respectively. In four dimensions, exactly one of the latter two is half-integer, so we can have either one or two square roots at the maximal cut, depending on the exponent of $\mathcal{B}_2$. Following the discussion in app.~\ref{app:gram}, we observe that
$\mathcal{B}_2$ and $\mathcal{E}_1$ are at most quadratic polynomials in the Baikov variables, while $\mathcal{B}_1$ is at most quartic, but for planar integral topologies has at most degree 2 in the individual variables. Using this, in the following we sort the 79 integral topologies according to their number of ISPs and the structure of the maximal cut of the respective corner integrals. Throughout this section, we use $P_{n}$ to denote a polynomial of total degree $n$.

\subsubsection*{Integrals with zero ISPs} 

There are $36$ integral topologies with zero ISPs:
\begin{equation*}
\twoloop[2]{1}{1}{1} \quad \enspace \twoloop[1]{2}{1}{1} \quad  \enspace \twoloop[2]{2}{1}{2} \quad  \enspace \twoloop[1]{3}{1}{1} \quad  \enspace \twoloop[2]{3}{1}{2} \quad  \enspace \twoloop[1]{4}{1}{1} \quad  \enspace \twoloop[2]{4}{1}{2}
\end{equation*}
\vspace{-0.8cm}
\begin{flalign*}
& \hspace{0.15cm} I_{1,1,1}^{**} \hspace{1cm} I_{2, 1, 1}^* \hspace{1.5cm} I_{2, 1, 2}^{**}  \hspace{1.55cm} I_{3, 1, 1}^* \hspace{1.3cm} I_{3, 1, 2}^{**} \hspace{1.6cm} I_{4, 1, 1}^* \hspace{1.5cm} I_{4, 1, 2}^{**} &&
\end{flalign*}
\vspace{-0.5cm}
\begin{equation*}
\twoloop[210]{4}{3}{4} \hspace{2cm} \twoloop[10]{5}{1}{1} \hspace{1.9cm} \twoloop[210]{5}{1}{2} \hspace{2.2cm} \twoloop[210]{5}{1}{3}
\end{equation*}
\vspace{-0.8cm}
\begin{flalign*}
& \hspace{0.4cm} \{I_{4, 3, 4},I_{4, 3, 4}^*,I_{4, 3, 4}^{**}\} \hspace{1.05cm} \{ I_{5, 1, 1},I_{5, 1, 1}^* \} \hspace{0.65cm} \{ I_{5, 1, 2},I_{5, 1, 2}^*,I_{5, 1, 2}^{**} \} \hspace{0.8cm} \{ I_{5, 1, 3},I_{5, 1, 3}^*,I_{5, 1, 3}^{**} \} &&
\end{flalign*}
\vspace{-0.5cm}
\begin{equation*}
\twoloop[210]{5}{1}{4} \hspace{2cm} \twoloop[210]{5}{1}{5} \hspace{2cm} \twoloop[210]{5}{2}{2} \hspace{2cm} \twoloop[210]{5}{2}{3}
\end{equation*}
\vspace{-0.8cm}
\begin{flalign*}
& \hspace{0.2cm} \{ I_{5, 1, 4},I_{5, 1, 4}^*,I_{5, 1, 4}^{**} \} \hspace{0.65cm} \{ I_{5, 1, 5},I_{5, 1, 5}^*,I_{5, 1, 5}^{**} \} \hspace{0.6cm} \{ I_{5, 2, 2},I_{5, 2, 2}^*,I_{5, 2, 2}^{**} \} \hspace{0.5cm} \{ I_{5, 2, 3},I_{5, 2, 3}^*,I_{5, 2, 3}^{**} \} &&
\end{flalign*}
\vspace{-0.5cm}
\begin{equation*}
\twoloop[210]{5}{2}{4} \hspace{2cm} \twoloop[210]{5}{3}{3}
\end{equation*}
\vspace{-0.75cm}
\begin{flalign*}
& \hspace{4.1cm} \{ I_{5, 2, 4},I_{5, 2, 4}^*,I_{5, 2, 4}^{**} \} \hspace{0.6cm} \{ I_{5, 3, 3},I_{5, 3, 3}^*,I_{5, 3, 3}^{**} \} &&
\end{flalign*}
Note that for three of the integral topologies, namely $I_{4,3,4}$ with any number of stars, the loop-by-loop Baikov representation does not simply follow from eq.~\eqref{eq:Baikov} since the propagators are not linearly independent; see sec.~\ref{sec:special_case_434} for a discussion of how to derive it. 

Since they have no ISPs, the integrals in these topologies automatically have algebraic leading singularities and thus admit a $d$log form on the maximal cut. This is trivially the case for the corner integral, where all propagators occur with unit power. In integrals with dots, the residue at the poles where the dotted propagators vanish leads to logarithmic derivatives of the Baikov polynomials, cf.\ eq.\ \eqref{eq:Baikov}, which are also algebraic on the maximal cut. Finally, since there are no ISPs in these topologies, the integrals can have no non-trivial numerator factors. Thus, our results hold for the entire sector.\footnote{From the IBP point of view, since there are no ISPs, one can generate only one master integral in the sector, which can always be chosen as the corner integral. As there is only one master integral, it trivially admits a $d$log form.}

\subsubsection*{Integrals with one ISP} 

There are $29$ integral topologies with one ISP. We will sort them into four types depending on the degree of the Baikov polynomials and their exponents, which turns out to be sufficient to determine their associated geometry.

\paragraph{Type 1.1:} 14 integral topologies for which the corner integral has maximal cut of the form $\int \frac{dz}{\sqrt{P_{2}(z)}}$ or $\int \frac{dz}{Q_{2}(z)\sqrt{P_{2}(z)}}$ :
\begin{equation*}
\twoloop[0]{2}{1}{1} \qquad \twoloop[1]{2}{1}{2} \qquad \twoloop[0]{4}{1}{1} \qquad \twoloop[10]{4}{1}{2}
\end{equation*}
\vspace{-0.8cm}
\begin{flalign*}
& \hspace{3.2cm} I_{2, 1, 1} \hspace{1.7cm} I_{2, 1, 2}^* \hspace{2cm} I_{4, 1, 1} \hspace{0.95cm} \{ I_{4, 1, 2},I_{4, 1, 2}^* \} &&
\end{flalign*}
\vspace{-0.4cm}
\begin{equation*}
\twoloop[210]{4}{1}{3} \hspace{2cm} \twoloop[210]{4}{1}{4} \hspace{2cm} \twoloop[210]{4}{2}{3}
\end{equation*}
\vspace{-0.7cm}
\begin{flalign*}
& \hspace{2.1cm} \{ I_{4, 1, 3}, I_{4, 1, 3}^*, I_{4, 1, 3}^{**} \} \hspace{0.55cm} \{ I_{4, 1, 4}, I_{4, 1, 4}^*, I_{4, 1, 4}^{**} \} \hspace{0.7cm} \{ I_{4, 2, 3}, I_{4, 2, 3}^*, I_{4, 2, 3}^{**} \} &&
\end{flalign*}
The square root $\sqrt{P_{2}(z)}$ does not define a non-trivial geometry, as can be seen from the configuration matrix for $y^2=z_0^2P_2(z/z_0)$, where we introduced $z_0$ to homogenize. Indeed, we can use the change of variables in eq.~\eqref{eq: change_of_variables_(z-r1)(z-r2)} to rationalize the square root $\sqrt{P_2(z)}$. Afterwards, we are able to take a residue, and indeed find an algebraic leading singularity. Note that it is not necessary to check further configuration matrices involving $Q_2$ or the third Baikov polynomial, which has vanishing exponent in four dimensions, since those intersections are zero-dimensional. Dots on the propagators again give logarithmic derivatives of the Baikov variables. While those can lead to new poles, the residues at those poles will still be algebraic. The effect of the logarithmic derivatives of the Baikov polynomials on the already existing poles is to change the value of the residues, but not the algebraic nature of the leading singularity. Similarly, numerators $\mathcal{N}$ can only change the residues at the poles, but not their algebraic nature. Thus, all integrals in these topologies admit a $d$log form on the maximal cut.

\paragraph{Type 1.2:} 10 integral topologies where the maximal cut of the corner integral takes the form $\int \frac{dz}{\sqrt{P_{4}(z)}}$ or $\int \frac{dz}{Q_{2}(z)\sqrt{P_{4}(z)}}$ :
\begin{equation*}
\twoloop[2]{2}{2}{2} \qquad \qquad \twoloop[210]{4}{2}{2} \qquad \qquad \twoloop[210]{4}{2}{4} \qquad \qquad \twoloop[210]{4}{3}{3}
\end{equation*}
\vspace{-0.85cm}
\begin{flalign*}
& \hspace{2.1cm} I_{2, 2, 2}^{**} \hspace{1.25cm} \{ I_{4, 2, 2}, I_{4, 2, 2}^*, I_{4, 2, 2}^{**} \} \hspace{0.3cm} \{ I_{4, 2, 4}, I_{4, 2, 4}^*, I_{4, 2, 4}^{**} \} \hspace{0.3cm} \{ I_{4, 3, 3}, I_{4, 3, 3}^*, I_{4, 3, 3}^{**} \} &&
\end{flalign*}
Since the corner integrals contain a polynomial of degree 4 underneath the square root at the maximal cut, they manifestly involve an elliptic curve -- as also confirmed from the configuration matrix perspective. 
We have checked that $P_4(z)$ has a non-zero discriminant for all of the integrals. As in the previous case, we do not need to check any configuration matrices including $Q_2$ or the third Baikov polynomial, since the corresponding intersections and hypersurfaces are zero-dimensional. Dots will again introduce logarithmic derivatives of the Baikov polynomials that can introduce new poles or change the order of already existing ones. However, all such integrands can be related to those of elliptic integrands of the first, second and third kind using integration-by-parts relations, see e.g.\ ref.\ \cite{Broedel:2017kkb}. Therefore, each of the integral topologies will be associated to an elliptic curve.

The first of these integral topologies was already known to be elliptic in generic kinematics~\cite{Lairez:2022zkj}, and had been particularly studied when one of the loops is massive~\cite{vonManteuffel:2017hms,Broedel:2019hyg,Becchetti:2025rrz,Coro:2025vgn}. The next 3 integral topologies ($I_{4, 2, 2}$ with any number of stars), which are known as the non-planar hexa-box, have been previously studied in massless kinematics, where they admit a $d$log form~\cite{Abreu:2018rcw, Chicherin:2018mue, Kardos:2022tpo}. In general kinematics, we find instead that they are elliptic; see the discussion in sec.~\ref{sec: special_case_422}. The remaining integral topologies also involve an elliptic curve, but they do not contribute in strictly four dimensions at two loops, as they are evanescent.

\paragraph{Type 1.3:} 4 integral topologies where the maximal cut of the corner integral takes the form $\int \frac{dz}{\sqrt{Q_{2}(z)}\sqrt{P_{2}(z)}}$ :
\begin{equation*}
\twoloop[0]{1}{1}{1} \quad \qquad \twoloop[0]{3}{1}{1} \quad \qquad \twoloop[1]{3}{1}{2} \quad \qquad \twoloop[2]{3}{1}{3}
\end{equation*}
\vspace{-0.9cm}
\begin{flalign*}
& \hspace{3cm} I_{1, 1, 1} \hspace{1.4cm} I_{3, 1, 1} \hspace{1.8cm} I_{3, 1, 2}^* \hspace{2.25cm} I_{3, 1, 3}^{**} &&
\end{flalign*}
As in the previous case, from the configuration matrix approach we expect these integrals to be elliptic on the maximal cut; recall eq.~\eqref{eq:config_matrix_sunrise} for the example of $I_{1,1,1}$. Indeed, for the corner integrals we find an elliptic curve after rationalizing either of the square roots using the change of variables from eq.~\eqref{eq: change_of_variables_(z-r1)(z-r2)}, with a defining polynomial that has non-zero discriminant in all cases. As in the previous case, dots and numerators do not change the elliptic nature of integrals, such that our analysis hold for the whole sector.

The integral topologies $I_{1, 1, 1}$ (the sunrise), $I_{3, 1, 1}$ and $I_{3, 1, 3}^{**}$ (the 4-point double box) were already known to be elliptic in general kinematics~\cite{Adams:2013nia,Adams:2014vja,Lairez:2022zkj,Doran:2023yzu}; see also refs.~\cite{Broadhurst:1993mw,Laporta:2004rb,Bloch:2013tra,Bonciani:2016qxi,Adams:2018kez,Adams:2018bsn,Muller:2022gec} for equal-mass limits. The remaining integral topology, $I_{3, 1, 2}^*$, was only known to be elliptic for specific kinematic values~\cite{Bonciani:2016qxi,Adams:2018kez}.

\paragraph{Type 1.4:} 1 integral topology where the maximal cut of the corner integral takes the form $\int \frac{dz}{\sqrt{Q_{2}(z)}\sqrt{P_{4}(z)}}$ :
\begin{equation*}
\twoloop[2]{3}{2}{2}
\end{equation*}
\vspace{-0.7cm}
\begin{flalign*}
& \hspace{7.2cm} I_{3, 2, 2}^{**} &&
\end{flalign*}
In this case, the sum of degrees of the polynomials is too large to satisfy the CY condition~\eqref{CY_condition_WP}, which indicates the presence of a geometry more complicated than a CY. In fact, simply combining the square roots we would obtain a polynomial of degree 6, which is associated to a genus-2 hyperelliptic curve; recall tab.~\ref{tab:geometries}. Indeed, for a specific mass configuration, the integrals in this non-planar double-box topology are known to be hyperelliptic~\cite{Huang:2013kh,Marzucca:2023gto}. As detailed in sec.~\ref{subsubsec:non-planardoublebox_4pt} for general kinematics, we can rationalize the square root of the quadratic polynomial, obtaining a square root over a polynomial of degree 8, which is associated to a genus-3 hyperelliptic curve. However, there exists an extra involution in the polynomial~\cite{Marzucca:2023gto}, which reduces the genus from 3 to 2 also for generic kinematics. 

Also here, we do not need to check any configuration matrices including the third Baikov polynomial, since the corresponding intersections and hypersurfaces are zero-dimensional. Dots will again introduce logarithmic derivatives of the Baikov polynomials that can introduce new poles or change the order of already existing ones, while the numerator $\mathcal{N}$ can cancel poles and change the values of the associated residues. However, all such integrands can be related to hyperelliptic integrands that are analogs of the elliptic integrands of first, second and third kind; see e.g.\ refs.\ \cite{Enriquez2021,Enriquez2022,DHoker:2023vax,DHoker:2024ozn,Baune:2024ber} as well as ref.~\cite{Duhr:2024uid} for an example. Therefore, each of the integral topologies will be associated to a genus-2 curve.

\subsubsection*{Integrals with two or three ISPs} 

There are $13$ integral topologies with two ISPs:
\begin{equation*}
\twoloop[0]{2}{1}{2} \hspace{1.75cm} \twoloop[0]{3}{1}{2} \hspace{1.75cm} \twoloop[1]{2}{2}{2} \hspace{1.75cm} \twoloop[10]{3}{1}{3}
\end{equation*}
\vspace{-0.9cm}
\begin{flalign*}
& \hspace{1.95cm} I_{2, 1, 2} \hspace{2.75cm} I_{3, 1, 2} \hspace{2.9cm} I_{2, 2, 2}^* \hspace{1.9cm} \{ I_{3, 1, 3}, I_{3, 1, 3}^* \} &&
\end{flalign*}
\vspace{-0.5cm}
\begin{equation*}
\twoloop[10]{3}{2}{2} \hspace{2cm} \twoloop[210]{3}{2}{3} \hspace{2cm} \twoloop[210]{3}{3}{3}
\end{equation*}
\vspace{-0.7cm}
\begin{flalign*}
& \hspace{2.7cm} \{ I_{3, 2, 2}, I_{3, 2, 2}^*\} \hspace{0.9cm} \{ I_{3, 2, 3}, I_{3, 2, 3}^*, I_{3, 2, 3}^{**} \} \hspace{0.7cm} \{ I_{3, 3, 3}, I_{3, 3, 3}^*, I_{3, 3, 3}^{**} \} &&
\end{flalign*}
which can a priori involve at most a two-dimensional geometry. We will look at each of these cases individually in sec.~\ref{sec:special_cases}. To tease some of the new results, we find that all of the integrals in the topologies above are elliptic or worse at the maximal cut. For example, we find that the 5- and 6-point non-planar double-box integral topologies ($I_{3, 2, 2}$ and $I_{3, 2, 2}^*$) involve a hyperelliptic curve of genus 3, in this case without an extra involution reducing the genus (as opposed to the 4-point case discussed above); see sec.~\ref{subsubsec:non-planardoublebox} for details. For the integral topology $I_{3, 3, 3}$, we obtain a polynomial of total degree 4 underneath the square root, and which is quartic in both variables. This defines a Del Pezzo surface of degree 2, a particular kind of Fano variety. Such a surface is known to be rationalizable~\cite{Schicho2005}, and the leading singularity of the corner integral becomes associated to a curve (not necessarily hyperelliptic) of geometric genus 3; see sec.~\ref{sec:special_case_goomba} and app.~\ref{app:rationalization_Del_Pezzo} for details.

Lastly, there is $1$ integral topology with three ISPs, known as the 5-point tardigrade integral~\cite{Bourjaily:2018yfy},
\begin{equation*}
\twoloop[0]{2}{2}{2}
\end{equation*}
\vspace{-0.7cm}
\begin{flalign*}
& \hspace{7.25cm} I_{2, 2, 2} &&
\end{flalign*}
which can at most depend on a three-dimensional geometry. For both this integral topology and its 4-point incarnation ($I_{2,2,2}^*$ above), we find that they involve a K3 surface on the maximal cut; see secs.~\ref{subsubsec:tardigrade} and~\ref{subsubsec:tardigrade_4pt} for details, respectively. In the 5-point case, this was already identified in ref.~\cite{Doran:2023yzu} for generic kinematics; see also refs.~\cite{Bourjaily:2018yfy,Lairez:2022zkj}.

\section{Full classification at two loops: Special cases}
\label{sec:special_cases}

As explained in the previous section, the structure of the Baikov polynomials for two-loop Feynman integrals on the maximal cut bounds the complexity of the associated geometry, and this bound is actually saturated in the majority of the integrals. However, there are some cases for which one needs to perform specific changes of variables to make the underlying geometry manifest. In this section, we will take a closer look at some elliptic integrals that are more challenging to analyze, as well as all cases involving a geometry beyond the elliptic curve. We organize them according to their complexity and resemblance, starting with the integral topologies containing an elliptic curve on the maximal cut, and moving to K3 surfaces and higher-genus curves. Lastly, we also include an example with an algebraic leading singularity due to a subtlety in deriving its loop-by-loop Baikov representation. Note that the specific parametrization used for each integral and its Baikov representation, along with the full analysis of the leading singularity, can be found in the \texttt{Mathematica} notebooks provided in the ancillary files.

Throughout this section, we will explicitly keep the Baikov polynomials which have vanishing exponents in integer dimensions. This is because they can introduce non-trivial geometries through higher orders in $\varepsilon$ or via master integrals other than the corner integral, so they must also be considered in the analysis; recall the discussion from sec.~\ref{sec: LBL_Baikov}. Therefore, our analysis of geometries also holds for any master integral in the sector, including any number of dots and ISPs. 

\subsection[5-point triangle-box]{\texorpdfstring{$\text{I}_{\text{3,1,2}}$}{I312}: 5-point triangle-box}
\label{subsubsec:5Tbox}

To begin with, let us consider the integral topology $I_{3,1,2}$, which is known as the 5-point triangle-box. This first example will allow us to illustrate our methodology, as well as to derive general results that apply to other topologies. 

This topology has two ISPs, $z_1$ and $z_2$, and in $D=4$ the corner integral takes the following form on the maximal cut, where we drop constant prefactors:
\begin{equation}
\twoloop[0]{3}{1}{2} \, \Bigg|_{\text{max-cut}} \propto \int \frac{dz_1 dz_2}{\sqrt{\mathcal{E}_1} \ \mathcal{B}_1^0 \ \mathcal{B}_2} \, . 
\end{equation}
The Baikov polynomials are schematically given by
\begin{equation}
\mathcal{E}_1= -\lambda(p_1^2,z_1,z_2) \, ,\qquad
\mathcal{B}_1=\sum\limits_{i,j=0}^{i+j\leq 2} \alpha_{i,j} \, z_{1}^{i} z_{2}^{j} \, , \qquad \mathcal{B}_2=\sum\limits_{i,j=0}^{i+j\leq 2} \beta_{i,j} \, z_{1}^{i} z_{2}^{j} \, ,
\end{equation}
where $\lambda$ denotes the Källén function, defined as 
\begin{equation}
\label{eq:Kallen}
\lambda(a,b,c) = a^2 + b^2 + c^2 - 2 ab - 2ac - 2bc\,.
\end{equation}

Since this integral topology has two ISPs, following the general procedure described in sec.\ \ref{sec: CICY_geo} we first need to check for the occurrence of a non-trivial two-dimensional geometry. 
However, since $\mathcal{E}_1$ is quadratic, its degree is too low to define a non-trivial geometry, as can be seen from the configuration matrix 
\begin{equation}
\label{eq:confmat11112}
[\mathbb{WP}^{1,1,1,1} \, || \, 2 ]
\end{equation}
in weighted projective space $[z_0:z_1:z_2:y] \in \mathbb{WP}^{1,1,1,1}$, where $z_0$ is the homogenization variable and $y$ denotes the square root.
Indeed, we can rationalize this square root, as we explicitly show below; recall also the discussion around eq.~\eqref{eq:rationalization_quadratic_polynomials}.

Proceeding to check for occurrences of non-trivial one-dimensional geometries, we first consider the configuration matrix for the corner integral in strict four dimensions, i.e.~when the Baikov polynomial $\mathcal{B}_1$ is absent but the other two are present. In this case, we have 
\begin{equation}
[\mathbb{WP}^{1,1,1,1} \, || \, 2 \ 2] \,.
\end{equation} 
As can be seen, it satisfies the CY condition~\eqref{CY_condition_WP}; thus, our analysis predicts that this integral involves an elliptic curve on the maximal cut, in agreement with ref.~\cite{Doran:2023yzu}. We can make the elliptic curve manifest through variable transformations. First, we can perform the $\text{GL}(2,\mathbb{C})$ transformation
\begin{equation}
\label{eq: GL2_transformation_kite}
z_{1,2} = \frac{1}{2} \left(\frac{z_{+}+p_1^4}{2p_1^2}\pm z_{-}\right)
\end{equation}
to bring the square root $\sqrt{\mathcal{E}_1}$ into the form $\sqrt{z_+ - z_-^2}$. Then, we can use the change of variables from eq.~\eqref{eq: change_of_variables_(z-c_squared)} to rationalize it by changing variables from $z_+$ to $x$. Afterwards, since there are no square roots remaining, we can safely take a residue in $z_-$. At the end, we obtain
\begin{equation}
\LS \left( I_{3,1,2} \right) = \frac{\varepsilon}{4 \pi ^3 } \int \frac{dx}{\sqrt{P_4(x)}} \, ,
\end{equation}
where $P_4(x)$ is a quartic polynomial with explicit dependence on $i$ in its (odd-powered) coefficients. Thus, changing $x \to i \, x$, the coefficients in the polynomial become real, and they correspond to rational functions of the kinematics. Note that this final transformation introduces an overall prefactor of $i$ through the Jacobian, so that the leading singularity only changes sign under the action of the Galois symmetry $i\rightarrow-i$; recall the discussion below eq.~\eqref{eq: change_of_variables_(z-c_squared)}. Hence, the choice of $i$ corresponds to two different realizations of the same, unique geometry. In particular, $P_4(x)$ is a quartic polynomial with non-vanishing discriminant, manifesting that the associated geometry is an elliptic curve. Note that in ref.~\cite{Becchetti:2025qlu}, a specific kinematic limit of this diagram was studied for the $q\bar{q}\rightarrow t\bar{t}W$ process. We have checked that under these kinematics, our elliptic curve has the same $j$-invariant as the one identified in ref.~\cite{Becchetti:2025qlu} after exchanging $p_1\leftrightarrow p_4$ to match conventions.

Next, we consider the contributions from the polynomial $\mathcal{B}_1$ to the configuration matrix, which becomes relevant at higher orders in $\varepsilon$ and for other master integrals in the sector, such as when placing a dot in one of the propagators of the triangle loop. The configuration matrix associated to $\mathcal{E}_1$ and $\mathcal{B}_1$ is also
\begin{equation}
[\mathbb{WP}^{1,1,1,1} \, || \, 2 \ 2]\,,
\end{equation}
which a priori indicates another elliptic curve. Rationalizing $\mathcal{E}_1$ and taking residues, the associated integral similarly becomes
\begin{equation}
\int \frac{dx}{\sqrt{Q_4(x)}} \, ,
\end{equation}
which gives an elliptic curve $y^2=Q_4(x)$. However, in this case $Q_4(x)$ has vanishing discriminant, as it is the perfect square of a polynomial of degree 2. Thus, the elliptic curve is degenerate, and results in an algebraic leading singularity. In fact, as will become apparent in the remaining examples in this section, we observe that the geometries defined by $\mathcal{E}_1$ and $\mathcal{B}_1$ are always degenerate. Indeed, this is a direct consequence of the Desnanot--Jacobi identity from eq.~\eqref{eq: Desnanot_Jacobi}. In particular, following eq.~\eqref{eq: degeneration_Desnanot_Jacobi}, the condition $\mathcal{B}_1=0$ imposes that $\mathcal{E}_1$ becomes a perfect square, explaining this degeneration.

Finally, note that we do not need to consider the configuration matrix involving $\mathcal{B}_1$ and $\mathcal{B}_2$ since this intersection is zero dimensional. Thus, together with the discussion around eq.~\eqref{eq:rationalization_quadratic_polynomials}, the cases above cover all non-trivial effects that dots and numerators can have on the leading singularity. All other inclusions of dots and numerators can change the values of the leading singularities, but not their geometric nature. Our results thus hold for all integrals in this sector. 

\subsection[5- and 6-point double box]{\texorpdfstring{$\text{I}_{\text{3,1,3}}^{*}$}{I313*} and \texorpdfstring{$\text{I}_{\text{3,1,3}}$}{I313}: 5- and 6-point double box}
\label{subsubsec:doublebox}

Next, we can study the integral topologies $I_{3,1,3}^{*}$ and $I_{3,1,3}$, which respectively correspond to the 5- and 6-point double box. They have two ISPs, denoted as $z_1$ and $z_2$. On the maximal cut in $D=4$, the corner integrals take the following form:
\begin{equation}\twoloop[10]{3}{1}{3} \, \Bigg|_{\text{max-cut}} \propto \int \frac{dz_1 dz_2}{\mathcal{E}_1^0 \sqrt{\mathcal{B}_1} \ \mathcal{B}_2} \, ,
\end{equation}
where the Baikov polynomials schematically are
\begin{equation}
\mathcal{E}_1=\sum\limits_{i,j=0}^{i+j\leq 2} \alpha_{i,j} \, z_{1}^{i} z_{2}^{j} \, , \qquad \mathcal{B}_{1}=\sum\limits_{i,j=0}^{i+j \leq 2} \beta_{i,j} \, z_{1}^{i} z_{2}^{j} \, , \qquad \mathcal{B}_2= \sum\limits_{i,j=0}^{i+j \leq 2} \gamma_{i,j} \, z_{1}^{i} z_{2}^{j} \, .
\end{equation}

In both cases, the structure of these polynomials is the same as in the previous example for the integral topology $I_{3,1,2}$, with the role of $\mathcal{E}_1$ and $\mathcal{B}_1$ interchanged. 
As in the integral topology $I_{3,1,2}$, no non-trivial two-dimensional geometry occurs, as can be seen from the configuration matrix for $\mathcal{B}_1$, which is identical to the one in eq.\ \eqref{eq:confmat11112}. 
Proceeding to check for the occurrence of non-trivial one-dimensional geometries, we consider the configuration matrix of the corner integral in strictly four dimensions. We obtain
\begin{equation}
[\mathbb{WP}^{1,1,1,1} \, || \, 2 \ 2]
\end{equation}
in weighted projective space $[z_0:z_1:z_2:y] \in \mathbb{WP}^{1,1,1,1}$, which satisfies the CY condition~\eqref{CY_condition_WP} for an elliptic curve. Note that the 6-point case was already known to be elliptic on the maximal cut in generic kinematics~\cite{Bloch:2021hzs,Lairez:2022zkj,Doran:2023yzu}; see also refs.~\cite{Caron-Huot:2012awx,Bourjaily:2017bsb,Vergu:2020uur,Kristensson:2021ani,Frellesvig:2021vdl,Morales:2022csr} for studies in the massless limit, where it is still elliptic.

Using changes of variables, we can again make the elliptic curve manifest for both topologies. First, we can perform a linear shift of the form $z_2\rightarrow a_0\, z_1 + z_2$, with $a_0$ chosen to reduce $\mathcal{B}_2$ into a linear polynomial in $z_1$. Then, using the transformation from eq.~\eqref{eq: change_of_variables_(z-r1)(z-r2)} to change from $z_1$ to $x_1$, we can rationalize the square root, and take a subsequent residue in $x_1$. For example, for the case of the 5-point corner integral, we obtain
\begin{equation}
\LS \left( I^{*}_{3,1,3} \right) = - \frac{\varepsilon \, P_3(p_i^2,s_{ij},s_{ijk}) \, \sqrt{Q_4(p_i^2,s_{ij},s_{ijk})}}{4 \sqrt{2} \, \pi^4} \int \frac{dz_2}{\sqrt{P_4(z_2)}} \, .
\end{equation} 
Here, $P_4(z_2)$ denotes a polynomial of degree 4 in $z_2$, and $P_3$, $Q_4$ are cubic and quartic polynomials in the kinematic variables, respectively. As a consequence, the leading singularity indeed corresponds to an integral over a smooth elliptic curve, as $P_4$ has non-zero discriminant, and the same occurs for the 6-point topology. Let us highlight that, while $P_4(z_2)$ has coefficients that are algebraic functions of the kinematics (they depend on $\sqrt{Q_4}$), the associated $j$-invariant is square-root free. Hence, there is a unique elliptic curve under the action of the Galois symmetry, which flips the sign of $\sqrt{Q_4}$. The same is true for all of the elliptic cases that follow in this paper.

Similarly to the case of sec.~\ref{subsubsec:5Tbox}, here there is also an apparent elliptic curve associated to $\mathcal{E}_1$ and $\mathcal{B}_1$, which we obtain by rationalizing $\mathcal{B}_1$ in the same way as for the corner integral. However, the defining polynomial of the curve has vanishing discriminant, as expected from eq.~\eqref{eq: degeneration_Desnanot_Jacobi}; thus, it degenerates into a rational curve and we have an algebraic leading singularity in the absence of other polynomials in the denominator. The intersection of $\mathcal{E}_1$ and $\mathcal{B}_2$ is zero dimensional.

Together with the discussion around eq.~\eqref{eq:rationalization_quadratic_polynomials}, the cases above cover all non-trivial effects that dots and numerators can have on the leading singularity. All other inclusions of dots and numerators can change the values of the leading singularities, but not their geometric nature. Our results thus hold for all integrals in these sectors.

\subsection[4-point kite]{\texorpdfstring{$\text{I}_{\text{2,1,2}}$}{I212}: 4-point kite}
\label{subsubsec:kite}

The integral topology $I_{2,1,2}$, known as the 4-point kite or slashed-box, has two ISPs, $z_1$ and $z_2$. In $D=4$, the maximal cut of the corner integral is given by
\begin{equation}
\twoloop[0]{2}{1}{2} \, \Bigg|_{\text{max-cut}} \propto \int \frac{dz_1 dz_2}{\sqrt{\mathcal{E}_1} \ \mathcal{B}_1^0 \sqrt{\mathcal{B}_2}} \, ,
\end{equation}
with the Baikov polynomials
\begin{equation}
\mathcal{E}_1= -\lambda(p_1^2,z_1,z_2) \, ,\qquad
\mathcal{B}_1=\sum\limits_{i,j=0}^{i+j\leq 2} \alpha_{i,j} \, z_{1}^{i} z_{2}^{j} \,, \qquad \mathcal{B}_2=\sum\limits_{i,j=0}^{i+j\leq 2} \beta_{i,j} \, z_{1}^{i} z_{2}^{j} \, ,
\end{equation}
where $\lambda$ is again the Källén function \eqref{eq:Kallen}.

In this case, the configuration matrix associated to the corner integral in strictly four dimensions, i.e.\ with $\mathcal{E}_1$ and $\mathcal{B}_2$, is
\begin{equation}
[\mathbb{WP}^{1,1,1,1,1} \, || \, 2 \ 2]
\end{equation}
in weighted projective space $[z_0:z_1:z_2:y_1:y_2] \in \mathbb{WP}^{1,1,1,1,1}$. We see that this does not define a non-trivial two-dimensional geometry since the sum of degrees is too low to satisfy the CY condition~\eqref{CY_condition_WP}. In fact, we can first rationalize $\sqrt{\mathcal{E}_1}$ using the same transformations as in sec.~\ref{subsubsec:5Tbox}. Then, the remaining square root is of degree 6, but is only quadratic in $z_-$. Therefore, we can rationalize it and take a residue, which leads to a square root over a quadratic polynomial in one variable. Rationalizing again, we finally obtain an algebraic leading singularity.

Proceeding to check for the occurrence of non-trivial one-dimensional geometries, we consider the configuration matrix associated to the polynomials $\mathcal{E}_1$ and $\mathcal{B}_1$, as well as $\mathcal{B}_1$ and $\mathcal{B}_2$. In both cases, the configuration matrix becomes
\begin{equation}
[\mathbb{WP}^{1,1,1,1} \, || \, 2 \ 2]
\end{equation}
in weighted projective space $[z_0:z_1:z_2:y] \in \mathbb{WP}^{1,1,1,1}$. Now, these configuration matrices satisfy the CY condition~\eqref{CY_condition_WP}, and define two elliptic curves. Just as in the previous examples, the elliptic curve associated with the polynomials $\mathcal{E}_1$ and $\mathcal{B}_1$, made manifest by the same rationalization as for the corner integral, is defined by a quartic polynomial with vanishing discriminant, as explained through eq.~\eqref{eq: degeneration_Desnanot_Jacobi}. Thus, it degenerates into a rational curve, which yields an algebraic leading singularity. By contrast, the polynomials $\mathcal{B}_1$ and $\mathcal{B}_2$ actually yield, after rationalizing $\mathcal{B}_1$ using the change of variables in eq.~\eqref{eq: change_of_variables_(z-r1)(z-r2)}, a smooth elliptic curve with non-zero discriminant:
\begin{equation}
\int\frac{dx}{\sqrt{P_4(x)}} \, .
\end{equation}

Hence, even though there would naively seem to be no non-trivial geometry associated to the corner integral in four dimensions, the Baikov polynomial with vanishing exponent actually introduces an elliptic curve in the integral topology through the combination of $\mathcal{B}_1$ and $\mathcal{B}_2$, which we would have missed otherwise. This is in complete agreement with ref.~\cite{Adams:2018kez}, where a particular equal-mass case was studied, showing that an elliptic curve explicitly appears in the leading singularity of the master integral with a dot in one propagator; see also refs.~\cite{Lairez:2022zkj,Doran:2023yzu,Bree:2025tug} for related studies.

The cases above cover all non-trivial effects that dots and numerators can have on the leading singularity. All other inclusions of dots and numerators can change the values of the leading singularities, but not their geometric nature. Our results thus hold for all integrals in this sector.

\subsection[Non-planar hexa-box]{\texorpdfstring{$\text{I}_{\text{4,2,2}}$}{I422} and starred versions: Non-planar hexa-box}
\label{sec: special_case_422}

Let us now consider the integral topologies $I_{4,2,2}$, $I_{4,2,2}^*$ and $I_{4,2,2}^{**}$, known as the non-planar hexa-box~\cite{Abreu:2018rcw, Chicherin:2018mue, Kardos:2022tpo}. Their loop-by-loop Baikov parametrization has one ISP, $z$, and in $D=4$, the maximal cut of the corner integrals is
\begin{equation}
\label{eq:I_422_scalar}
\twoloop[210]{4}{2}{2} \, \Bigg|_{\text{max-cut}} \propto \int \frac{dz}{\mathcal{E}_1^0 \sqrt{\mathcal{B}_1} \ \mathcal{B}_2} \, ,
\end{equation}
where
\begin{equation}
\mathcal{E}_1= \sum\limits_{i=0}^{2} \alpha_{i} \, z^{i} \, , \qquad \mathcal{B}_1= \sum\limits_{i=0}^{4} \beta_{i} \, z^{i} \, , \qquad \mathcal{B}_2=\sum\limits_{i=0}^{2} \gamma_{i} \, z^{i} \, .
\end{equation}

Since there is only one ISP, we only need to check for non-trivial one-dimensional geometries, which can only be introduced through $\sqrt{\mathcal{B}_1}$. 
The configuration matrix of the square root $\sqrt{\mathcal{B}_1}$ is given by 
\begin{equation}
[\mathbb{WP}^{1,1,2} \, || \, 4]
\end{equation}
 in weighted projective space $[z_0:z:y] \in \mathbb{WP}^{1,1,2}$, which satisfies the CY condition~\eqref{CY_condition_WP} for an elliptic curve.
Indeed, $\mathcal{B}_1$ has non-vanishing discriminant, confirming the occurrence of this elliptic geometry. To manifest the form \eqref{eq:geodef}, we can choose a different master integral,
\begin{equation}
\twoloop[210]{4}{2}{2} \, \times \det G(k_2,p_1,p_2,p_3,p_4) \, \Bigg|_{\text{max-cut}} \propto \int \frac{dz}{\sqrt{\mathcal{B}_1}}\,,
\end{equation}
with a numerator that precisely cancels the $\mathcal{B}_2$ appearing in eq.~\eqref{eq:I_422_scalar}, where $k_2$ is the second loop momenta in the integration order. These integrands no longer have simple poles, but only contain the branch cuts of the elliptic curve. Thus, these integral topologies depend on an elliptic curve at the maximal cut.\footnote{From the perspective of loop-momentum representation, this argument means that the elliptic structure presumably starts to contribute at $\mathcal{O}(\varepsilon^1)$, while the finite part admits a $d$log form. This is because the loop momenta can have more than 8 degrees of freedom at higher orders in $\varepsilon$, so the maximal cut does not completely localize it.} Let us note that this case is completely analogous to elliptic integrals of the third kind; recall the discussion in sec.~\ref{sec: identifying_geometries}.

Together with the discussion around eq.~\eqref{eq:rationalization_quadratic_polynomials}, the cases above cover all non-trivial effects that dots and numerators can have on the leading singularity. All other inclusions of dots and numerators can change the values of the leading singularities, but not their geometric nature. Our results thus hold for all integrals in these sectors.

Different massless versions of this integral topology have previously been investigated in refs.~\cite{Abreu:2018rcw, Kardos:2022tpo, Chicherin:2018mue}. In those cases, the integrals admit a $d$log form, but the $\mathcal{B}_2$ factor in the denominator is still present, and in ref.~\cite{Chicherin:2018mue} a canonical basis was picked that explicitly cancels this factor with a corresponding numerator as discussed above.

\subsection[5-point tardigrade]{\texorpdfstring{$\text{I}_{\text{2,2,2}}$}{I222}: 5-point tardigrade}
\label{subsubsec:tardigrade}

The integral topology $I_{2,2,2}$, known as the 5-point tardigrade, has three ISPs, $z_1$, $z_2$ and $z_3$. In $D=4$, the maximal cut of the corner integral takes the form
\begin{equation}\twoloop[0]{2}{2}{2} \, \Bigg|_{\text{max-cut}} \propto \int \frac{dz_1 dz_2 dz_3}{\mathcal{E}_1^0 \sqrt{\mathcal{B}_1} \ \mathcal{B}_2} \, , 
\end{equation}
with the Baikov polynomials
\begin{equation}
\mathcal{E}_1 = \!\! \sum\limits_{i,j,k=0}^{i+j+k \leq 2} \!\! \alpha_{i,j,k} \, z_{1}^{i} z_{2}^{j} z_{3}^{k} \,, \quad\;
\mathcal{B}_1 = \!\! \sum\limits_{i,j,k=0}^{i+j+k \leq 4} \!\! \beta_{i,j,k}  \, z_{1}^{i} z_{2}^{j} z_{3}^{k} \,, \quad\;
\mathcal{B}_2 = \!\! \sum\limits_{i,j,k=0}^{i+j+k \leq 2} \!\! \gamma_{i,j,k} \, z_{1}^{i} z_{2}^{j} z_{3}^{k} \,.
\end{equation}

Since we have three ISPs, we should first check the presence of a three-dimensional geometry, which could only arise from $\sqrt{\mathcal{B}_1}$. However, one can realize that the degree-4 terms in $\mathcal{B}_1$ factorize as $z_2^2 (z_1+z_2+z_3)^2$, and that the same factorization occurs for the degree-3 terms, i.e.\ they factorize as $z_2(z_1+z_2+z_3)P_1(z_1,z_2,z_3)$, where $P_1$ is a linear polynomial. Hence, we can perform a shift $z_1 \rightarrow z_1-z_2-z_3$ to reduce $\mathcal{B}_1$ to a polynomial of overall degree 4 but of degree 2 in each integration variable. Then, following the argument in sec.~\ref{sec: LBL_Baikov}, we can rationalize it. Consequently, we obtain an algebraic leading singularity if no other polynomials are present, e.g.\ because $\mathcal{B}_2$ has been canceled by a corresponding numerator.

Therefore, we move on to two-dimensional geometries. First, as already discussed in eqs.~\eqref{eq:tardigrade_example}--\eqref{eq:tardigrade_example_config_matr}, the corner integral involves a K3 surface, which can be seen from its configuration matrix. In order to make this geometry manifest, one can first shift $z_1 \rightarrow z_1-z_2-z_3$ as discussed above. With a subsequent shift $z_2\rightarrow z_2 + c \, z_3$, we can thus eliminate the $z_3^2$ term in $\mathcal{B}_2$ by solving for $c$. Afterwards, we can use the transformation in eq.~\eqref{eq: change_of_variables_(z-r1)(z-r2)} to rationalize $\mathcal{B}_1$ with respect to $z_3$ and take a residue, which yields
\begin{equation}
\LS \left( I^{}_{2,2,2} \right) = \frac{\varepsilon \, \sqrt{P_4(p_i^2,s_{ij},s_{ijk})} }{16\pi^4 \, P_3(p_i^2,m_i^2,s_{ij},s_{ijk})} \int \frac{dz_1 dz_2}{\sqrt{P_{6}(z_1,z_2)}} \, ,
\end{equation}
where the $P_n$ are polynomials of overall degree $n$. The polynomial $P_{6}(z_1,z_2)$ is of degree 6 and 4 in the individual variables $z_1$ and $z_2$, respectively, and of overall degree 6. Since in addition it has non-zero discriminant with respect to both variables, the leading singularity thus explicitly defines an integral over a K3 surface; cf.~sec.~\ref{sec: identifying_geometries}.

This integral was already studied in ref.~\cite{Doran:2023yzu} for generic masses; see also refs.~\cite{Bourjaily:2018yfy,Lairez:2022zkj}. There, starting from the Schwinger parametrization, the authors find that the integral is associated to a conic fibration whose critical locus defines a K3 surface, with 6 singularities of type A1.\footnote{Effectively, this means that this K3 surface is (locally away from the singular points) isomorphic to a different surface where all singular points have been unfolded into circles; see e.g.\ ref.~\cite{Slodowy:1983} for a mathematical review. Importantly, this implies that the surface does not degenerate.} Furthermore, using the code given in app.~B of ref.~\cite{Doran:2023yzu}, it was found that this surface has Picard rank 11 at a generic numerical point, which further characterizes this geometry.

Looking at the contributions from $\mathcal{E}_1$, we observe that in combination with $\mathcal{B}_1$ it also defines an apparent K3 surface. Concretely, following the same procedure to rationalize $\mathcal{B}_1$ as in the corner integral, we obtain
\begin{equation}
\int \frac{dz_1 dz_2}{\sqrt{Q_6(z_1,z_2)}} \, .
\end{equation}
However, the corresponding polynomial $Q_6(z_1,z_2)$ has vanishing discriminant with respect to both variables, which is expected from the Desnanot--Jacobi identity and eq.~\eqref{eq: degeneration_Desnanot_Jacobi}. Hence, it does not define a non-trivial two-dimensional geometry. 
Together with the discussion around eq.~\eqref{eq:rationalization_quadratic_polynomials}, the cases above cover all non-trivial effects that dots and numerators can have on the leading singularity, such that our results hold for all integrals in this sector.

\subsection[4-point tardigrade]{\texorpdfstring{$\text{I}_{\text{2,2,2}}^{*}$}{I222*}: 4-point tardigrade}
\label{subsubsec:tardigrade_4pt}

Next, we can consider the integral topology $I_{2,2,2}^{*}$, which corresponds to the 4-point tardigrade. This integral topology has instead two ISPs, $z_1$ and $z_2$. In $D=4$, the maximal cut for the corner integral is given by
\begin{equation}
\label{eq: maxcut_4pt_tardigrade}
\twoloop[1]{2}{2}{2} \, \Bigg|_{\text{max-cut}} \propto \int \frac{dz_1 dz_2}{\mathcal{E}_1^0 \sqrt{\mathcal{B}_1} \sqrt{\mathcal{B}_2}} \, , 
\end{equation}
with
\begin{equation}
\mathcal{E}_1= \sum\limits_{i,j=0}^{i+j \leq 2} \alpha_{i,j} \, z_{1}^{i} z_{2}^{j} \, , \qquad \mathcal{B}_1= \sum\limits_{i,j=0}^{i+j \leq 4} \beta_{i,j} \, z_{1}^{i} z_{2}^{j} \, , \qquad \mathcal{B}_2=\sum\limits_{i,j=0}^{i+j\leq 2} \gamma_{i,j} \, z_{1}^{i} z_{2}^{j} \, .
\end{equation}

Here, a two-dimensional geometry could in principle arise for the corner integral in strictly four dimensions, i.e.\ for $\sqrt{\mathcal{B}_1}$ and $\sqrt{\mathcal{B}_2}$. The corresponding configuration matrix is given by
\begin{equation}
[\mathbb{WP}^{1,1,1,1,2} \, || \, 2 \ 4]
\end{equation}
in weighted projective space $[z_0:z_1:z_2:y_1:y_2] \in \mathbb{WP}^{1,1,1,1,2}$. Notably, it satisfies the CY condition~\eqref{CY_condition_WP} for a K3 surface. However, unlike for the 5-point tardigrade studied in the previous subsection, in this case we did not succeed in finding a change of variables that results in a single square root over a polynomial of degree 6, which would manifest the form \eqref{eq:geodef}. Joining the square roots in eq.~\eqref{eq: maxcut_4pt_tardigrade} yields such a result, but this slightly changes the geometry; recall the discussion of the sunrise integral in sec.\ \ref{sec: CICY_geo}, where joining two square roots gave an isogenous elliptic curve, as well as the discussion in the following subsection. 

To definitively prove that this integral topology involves a K3 surface, we follow the same procedure as in ref.~\cite{Doran:2023yzu}, see also sec.~\ref{subsubsec:tardigrade}, but now taking the corresponding soft limit.\footnote{In the conventions of ref.~\cite{Doran:2023yzu}, this is the limit $k\rightarrow0$.} In this case, we find again a K3 surface with 6 singularities of type A1. Applying also their code to this example, we were able to numerically derive that the Picard rank of this K3 surface is still generically 11.%
\footnote{We are grateful to Eric Pichon-Pharabod for enlightening communication on this code.}

As in the previous topologies, the discussion above covers all non-trivial effects that dots and numerators can have on the leading singularity, such that our results hold for all integrals in this sector.

\subsection[4-point non-planar double box]{\texorpdfstring{$\text{I}_{\text{3,2,2}}^{**}$}{I322**}: 4-point non-planar double box}
\label{subsubsec:non-planardoublebox_4pt}

Let us now consider the integral topology $I_{3,2,2}^{**}$, known as the 4-point non-planar double-box, which has one ISP, denoted as $z$. In $D=4$, the maximal cut of the corner integral is
\begin{equation}
\label{eq:322cut}
\twoloop[2]{3}{2}{2} \, \Bigg|_{\text{max-cut}} \propto \int \frac{dz}{\mathcal{E}_1^0 \sqrt{\mathcal{B}_1} \sqrt{\mathcal{B}_2}} \, ,
\end{equation}
where
\begin{equation}
\mathcal{E}_1= \sum\limits_{i=0}^{2} \alpha_{i} \, z^{i} \, , \qquad \mathcal{B}_1= \sum\limits_{i=0}^{4} \beta_{i} \, z^{i} \, , \qquad \mathcal{B}_2=\sum\limits_{i=0}^{2} \gamma_{i} \, z^{i} \, .
\end{equation}

Since we have only one ISP, we only need to check for the presence of non-trivial one-dimensional geometries. The configuration matrix associated to $\mathcal{B}_1$ alone defines an elliptic curve with non-zero discriminant. By contrast, the configuration matrix associated to the corner integral in strict four dimensions (so, for the combination of $\mathcal{B}_1$ and $\mathcal{B}_2$) does not satisfy the CY condition~\eqref{CY_condition_WP} because the sum of degrees is greater than the sum of projective weights. Thus, it may signal the presence of a non-trivial one-dimensional geometry more complicated than an elliptic curve.

Taking the corner integral in strictly four dimensions, we can use the transformation in eq.~\eqref{eq: change_of_variables_(z-r1)(z-r2)} to rationalize $\mathcal{B}_2$, which directly results in
\begin{equation}
\label{eq:I322_genus_3}
\LS \left( I^{**}_{3,2,2} \right) = - \frac{4 \, P_2(p_i^2,s,t,u)^{3/2}}{\pi^4} \int \frac{x \, dx}{\sqrt{P_8(x)}} \, ,
\end{equation}
where $P_2$ is a quadratic polynomial in the kinematic variables. As can be seen, the leading singularity corresponds to an integral over the square root of a degree-8 polynomial with non-zero discriminant, which defines a hyperelliptic curve of genus 3. However, we need to be more careful, as it has been shown for a certain equal-mass limit that this integral actually involves a genus-2 curve in disguise~\cite{Marzucca:2023gto}. The origin of such a simplification is the presence of a so-called extra involution in the defining equation of the hyperelliptic curve. Let us briefly recap the definition of such an extra involution for the convenience of the reader.

A hyperelliptic curve of genus $g$ may be expressed by the equation 
\begin{align}\label{eq:defining_equation_hyperelliptic_g}
y^2 = \sum_{i=0}^{2g+2} C_i \, x^i\,,
\end{align}
where, importantly, the coefficients are sufficiently generic. An involution is a discrete symmetry of the defining equation, such as $y \rightarrow -y$, which clearly leaves eq.~\eqref{eq:defining_equation_hyperelliptic_g} invariant since $y$ appears only squared. Similarly, if the coefficients in eq.~\eqref{eq:defining_equation_hyperelliptic_g} satisfied $C_i = 0$ for all odd $i$, then $x$ would only appear with even powers, and the defining equation would have the extra involution $x \rightarrow -x$. In such a case, there exists a change of variables $x^2 \rightarrow w$, which would manifest the fact that the hyperelliptic curve is actually of a lower genus than is apparent from the degree of the polynomial. However, such an extra involution may not always be explicit in the defining equation, i.e.\ it can also exist in the presence of odd powers of $x$. This is because, for a given hyperelliptic curve, its corresponding equation is not unique: All equations related by Möbius transformations
\begin{equation}
x \rightarrow \frac{a x + b}{c x + d} \, , \qquad \qquad y\rightarrow \frac{y}{(c x + d)^{2g+2}}
\end{equation}
for $a,b,c,d\in \mathbb{Z}$ with $ad-bc\neq 0$ 
describe the same hyperelliptic curve. Hence, in order to rule out the possibility of an additional involution, we must check that eq.~\eqref{eq:defining_equation_hyperelliptic_g} cannot be related to an equation with an explicit extra involution. This can be done using the algorithm given in ref.~\cite{Marzucca:2023gto}. In the case of eq.~\eqref{eq:I322_genus_3}, there indeed exists a Möbius transformation that manifests such an extra involution. Therefore, the integral topology $I_{3,2,2}^{**}$ at the maximal cut actually depends on a hyperelliptic curve of genus $2$ instead of $3$, as identified for a particular equal-mass configuration in ref.~\cite{Marzucca:2023gto}.

Note that one would arrive at the same conclusion by naively combining the square roots in eq.\ \eqref{eq:322cut}. However, doing so slightly changes the geometry; recall the discussion of the sunrise integral in sec.\ \ref{sec: CICY_geo}, where joining two square roots results in an isogenous elliptic curve.  

As in the previous topologies, the cases above cover all non-trivial effects that dots and numerators can have on the leading singularity, such that our results hold for all integrals in this sector.

\subsection[5- and 6-point non-planar double box]{\texorpdfstring{$\text{I}_{\text{3,2,2}}^{*}$}{I322*} and \texorpdfstring{$\text{I}_{\text{3,2,2}}$}{I322}: 5- and 6-point non-planar double box}
\label{subsubsec:non-planardoublebox}

Similarly, let us now consider the integral topologies $I_{3,2,2}^{*}$ and $I_{3,2,2}$, which correspond to the 5- and 6-point non-planar double box, respectively. They have two ISPs, denoted as $z_1$ and $z_2$, and in $D=4$ the maximal cut of the corner integrals becomes
\begin{equation}\twoloop[10]{3}{2}{2} \, \Bigg|_{\text{max-cut}} \propto \int \frac{dz_1 dz_2}{\mathcal{E}_1^0 \sqrt{\mathcal{B}_1} \ \mathcal{B}_2} \, .
\end{equation}
For both integrals, the Baikov polynomials take the form
\begin{equation}
\mathcal{E}_1= \sum\limits_{i,j=0}^{i+j \leq 2} \alpha_{i,j} \, z_{1}^{i} z_{2}^{j} \, , \qquad \mathcal{B}_1= \sum\limits_{i,j=0}^{i+j \leq 4} \beta_{i,j} \, z_{1}^{i} z_{2}^{j} \, , \qquad \mathcal{B}_2=\sum\limits_{i,j=0}^{i+j\leq 2} \gamma_{i,j} \, z_{1}^{i} z_{2}^{j} \, .
\end{equation}
In the following, we study the 5-point topology for simplicity, but let us note that a similar procedure holds for the 6-point case.

Since we have two ISPs, we start by checking for non-trivial two-dimensional geometries, which can only arise from $\sqrt{\mathcal{B}_1}$. However, just as in the case of the 5-point tardigrade of sec.~\ref{subsubsec:tardigrade}, a shift $z_2 \to z_2 + z_1$ reduces the polynomial $\mathcal{B}_1$ to be of degree 4 but at most quadratic in both variables. This is because the degree-4 terms in $\mathcal{B}_1$ factorize as $z_1^2(z_1-z_2)^2$, and the same occurs for the degree-3 terms, i.e.\ they are given by $z_1(z_1-z_2)P_1(z_1,z_2)$, where $P_1$ is a linear polynomial. Therefore, by the argument in sec.~\ref{sec: LBL_Baikov}, we can fully rationalize $\sqrt{\mathcal{B}_1}$ and obtain an algebraic leading singularity if no other polynomials are present in the denominator.

As a consequence, we proceed to one-dimensional geometries. In this case, they could arise from the corner integral in strict four dimensions, i.e.\ in the presence of both $\sqrt{\mathcal{B}_1}$ and $\mathcal{B}_2$, which has configuration matrix
\begin{equation}
[\mathbb{WP}^{1,1,1,2} \, || \, 2 \ 4]\,.
\end{equation}
The sum of the degrees is too high to satisfy the CY condition~\eqref{CY_condition_WP}, indicating the possible occurrence of a more complicated one-dimensional geometry than an elliptic curve. First, we can shift $z_2\rightarrow z_2 + z_1$ as indicated above, which reduces $\mathcal{B}_{1}$ to a polynomial of degree 4 with at most degree 2 in both $z_1$ and $z_2$. Then, using the change of variables in eq.~\eqref{eq: change_of_variables_(z-r1)(z-r2)} from $z_2$ to $x_2$, and taking a subsequent residue in $x_2$ we obtain
\begin{equation}
\int \frac{dz_1}{\sqrt{P_2 (z_1)}\sqrt{P_4 (z_1) + \sqrt{P_2 (z_1)} \, P_3 (z_1)}} \, ,
\end{equation}
where the $P_n(z_1)$ are polynomials of degree $n$ in $z_1$. Hence, we can rationalize $\sqrt{P_2 (z_1)}$ with the same transformation, now changing from $z_1$ to $x$. Reintroducing the prefactors, we finally obtain
\begin{equation}
\LS \left( I^{*}_{3,2,2} \right) = \frac{i \, \varepsilon \, \sqrt{P_3(p_i^2,s_{ij},s_{ijk})} \, \sqrt{P_4(p_i^2,s_{ij},s_{ijk})} \, P_6(p_i^2,s_{ij},s_{ijk})^{7/4}}{2 \sqrt{2} \, \pi^4} \int \frac{x \, dx}{\sqrt{P_8(x)}} \, .
\end{equation} 
We find that the leading singularity corresponds to an integral over the square root of a degree-$8$ polynomial with non-vanishing discriminant, and hence defines an integral over a hyperelliptic curve of genus $3$. Unlike for the 4-point case, the hyperelliptic curves appearing in the leading singularities of the 5- and 6-point non-planar double-box integrals have no extra involutions. Thus, these integral topologies truly involve hyperelliptic curves of genus $3$ at the level of the maximal cut.

Next, we can take into account the contributions from $\mathcal{E}_1$, which has vanishing exponent. For both integral topologies, the combination of $\mathcal{E}_1$ with $\mathcal{B}_1$ yields, after rationalizing $\mathcal{B}_1$ as in the corner integral, an apparent genus-3 hyperelliptic curve. However, as expected from the discussion around eq.~\eqref{eq: degeneration_Desnanot_Jacobi}, its defining polynomial has vanishing discriminant and becomes a perfect square, resulting in an algebraic leading singularity. The intersection of $\mathcal{E}_1$ and $\mathcal{B}_2$ is zero dimensional.

As before, the cases above cover all non-trivial effects that dots and numerators can have on the leading singularity, thus our results hold for all integrals in these sectors.

\subsection[Non-planar double pentagon]{\texorpdfstring{$\text{I}_{\text{3,2,3}}$}{I323} and starred versions: Non-planar double pentagon}
\label{subsubsec:I323}

The integral topologies $I_{3,2,3}$, $I^*_{3,2,3}$ and $I^{**}_{3,2,3}$, known as the non-planar double pentagon~\cite{Chicherin:2018mue, Chicherin:2018old}, have two ISPs, $z_1$ and $z_2$. In $D=6$ dimensions, the corner integrals take the following form on the maximal cut:
\begin{equation}
\twoloop[210]{3}{2}{3} \, \Bigg|_{\text{max-cut}} \propto \int \frac{dz_1 dz_2}{\sqrt{\mathcal{E}_1} \ \mathcal{B}_1^0 \ \mathcal{B}_2^0}\,.
\end{equation}
In the three cases, the Baikov polynomials have the structure
\begin{equation}
\mathcal{E}_1= \sum\limits_{i,j=0}^{i+j \leq 2} \alpha_{i,j} \, z_{1}^{i} z_{2}^{j} \, , \qquad \mathcal{B}_1= \sum\limits_{i,j=0}^{i+j \leq 4} \beta_{i,j} \, z_{1}^{i} z_{2}^{j} \, , \qquad \mathcal{B}_2=\sum\limits_{i,j=0}^{i+j\leq 2} \gamma_{i,j} \, z_{1}^{i} z_{2}^{j} \, ,
\end{equation}
where the exponents of the latter two vanish in $D=6$. 

In this case, the configuration matrices for the corner integrals reveal that no non-trivial two-dimensional geometry occurs in strictly six dimensions. Indeed, for the three topologies, since $\mathcal{E}_1$ is a quadratic polynomial, we can rationalize it and obtain an algebraic leading singularity in the absence of further polynomials in the denominator; recall the discussion in sec.~\ref{sec: LBL_Baikov}.

We proceed by checking for non-trivial one-dimensional geometries. We start by the polynomial $\mathcal{B}_1$ alone, since it is a quartic polynomial that could introduce a non-trivial geometry. In this case, we have a configuration matrix
\begin{equation}
[\mathbb{WP}^{1,1,1} \, || \, 4]
\end{equation}
in weighted projective space $[z_0:z_1:z_2] \in \mathbb{WP}^{1,1,1}$, which does not satisfy the CY condition~\eqref{CY_condition_WP} because the degree of $\mathcal{B}_1$ is too high. Importantly, however, for the three integral topologies we are able to perform a shift $z_1 \to z_1 + c \, z_2$ such that $\mathcal{B}_1$ becomes only quadratic in $z_2$. Consequently, we can rationalize it with respect to $z_2$ and take a residue, which reveals a square root of a degree-6 polynomial in $z_1$ with non-zero discriminant,
\begin{equation}
\label{eq: genus-2_323}
\int\frac{dz_1}{\sqrt{P_6(z_1)}}\, .
\end{equation}
Therefore, it defines an integral over a hyperelliptic curve of genus 2.

Similarly, we can consider the polynomials $\mathcal{E}_1$ and $\mathcal{B}_2$. For all three topologies, the configuration matrix becomes
\begin{equation}
[\mathbb{WP}^{1,1,1,1} \, || \, 2 \ 2]
\end{equation}
in weighted projective space $[z_0:z_1:z_2:y] \in \mathbb{WP}^{1,1,1,1}$, which satisfies the CY condition~\eqref{CY_condition_WP} for an elliptic curve. However, these apparent elliptic curves, made manifest by rationalizing $\mathcal{E}_1$ as in the corner integral, have a defining polynomial with vanishing discriminant. Thus, they actually lead again to an algebraic leading singularity. In this case, the origin of the degeneracy is a different version of the Desnanot--Jacobi identity from eq.~\eqref{eq: Desnanot_Jacobi}. In the case of the three integral topologies, we can take $M=\mathcal{B}_2 \equiv \det G(k_2,p_1,p_2,p_3,p_4)$, where $p_i$ denote external momenta, and obtain the identity
\begin{equation}
\det G(p_1,p_2,p_3,p_4) \, \mathcal{E}_1 = (M^1_n)^2 + \det G(p_1,p_2,p_3) \, \mathcal{B}_2 \,,
\end{equation}
with $\mathcal{E}_1 \equiv \det G(k_2,p_1,p_2,p_3)$. Hence, the zero locus of $\mathcal{B}_2$ explains the perfect square found in $\mathcal{E}_1$.

Next, we consider the combination of $\mathcal{E}_1$ with $\mathcal{B}_1$. In such a case, after performing the changes of variables from eq.~\eqref{eq: change_of_variables_(z-r1)(z-r2)} and taking a residue, we are led to 
\begin{equation}
\label{eq: genus-3_323}
\int\frac{dx}{\sqrt{Q_8(x)}\sqrt{Q_6(x)+Q_3(x)\sqrt{Q_8(x)}}} \, .
\end{equation}
As can be seen, we obtain a nested square root together with a square root of a degree-8 polynomial, where the $Q_8(x)$ is the same in both places where it occurs. Following the argument around the Desnanot--Jacobi identity in eq.~\eqref{eq: degeneration_Desnanot_Jacobi}, one could naively think that the combination $Q_6(x)+Q_3(x)\sqrt{Q_8(x)}$ should eventually become a perfect square, just as in the previous examples. However, this expectation fails in this case because the topology is non-planar: the Gram determinant $\det G(k_1,\bar{p}_1)$ actually depends on the ISPs, which prevents $\mathcal{E}_1$ from being a perfect square in the zero locus of $\mathcal{B}_1$. A priori, since $Q_8(x)$ moreover has non-zero discriminant, this result would signal a hyperelliptic curve of genus 3. Nevertheless, the equation $y^2=Q_8(x)$ has an extra involution (recall the discussion in sec.~\ref{subsubsec:non-planardoublebox_4pt}) that reduces the genus from 3 to 2, in the three integral topologies given by the simple transformation $x \to (x'-2)/(x'+2)$. In fact, the resulting genus-2 hyperelliptic curve is isomorphic to the one previously obtained in eq.~\eqref{eq: genus-2_323}. This can be seen by comparing the so-called absolute Igusa invariants~\cite{Igusa:1960}, which uniquely characterize genus-2 curves and can be easily computed using the code from ref.~\cite{Marzucca:2023gto}. Thus, while we have not found an expression with a single square root when taking both polynomials $\mathcal{E}_1$ and $\mathcal{B}_1$ into account, our analysis is consistent with the presence of a genus-2 hyperelliptic curve in these integral topologies at the level of the maximal cut.

The cases above cover all non-trivial effects that dots and numerators can have, such that our results hold for all integrals in these sectors.

\subsection[Non-planar double hexagon]{\texorpdfstring{$\text{I}_{\text{3,3,3}}$}{I333} and starred versions: Non-planar double hexagon}
\label{sec:special_case_goomba}

The integral topologies $I_{3,3,3}$, $I_{3,3,3}^*$ and $I_{3,3,3}^{**}$, which correspond to the non-planar double hexagon -- also known as the goomba -- have two ISPs, denoted as $z_1$ and $z_2$. The corner integrals take the following form on the maximal cut in $D=6$ dimensions:
\begin{equation}
\twoloop[210]{3}{3}{3} \, \Bigg|_{\text{max-cut}} \propto \int \frac{dz_1 dz_2}{\sqrt{\mathcal{B}_1}} \, ,
\end{equation}
with the Baikov polynomials being
\begin{equation}
\mathcal{B}_1= \sum\limits_{i,j=0}^{i+j \leq 4} \alpha_{i,j} \, z_{1}^{i} z_{2}^{j} \, .
\label{eq:goombapoly}
\end{equation}

Crucially, for these integral topologies, the Baikov polynomials $\mathcal{E}_1$ and $\mathcal{B}_2$ cancel each other in any dimension, as they are both related to the Gram determinant 
\begin{equation}
\det G(k_2,p_1,p_2,p_3,p_4) = \sum\limits_{i,j=0}^{i+j \leq 2} \beta_{i,j} \, z_{1}^{i} z_{2}^{j} \, ,
\end{equation}
but with opposite exponents. Hence, we actually only have one Baikov polynomial. The corresponding configuration matrix shows that the degree is too low to satisfy the CY condition~\eqref{CY_condition_WP}, indicating the presence of a rationalizable geometry.

In this case, since the polynomial $\mathcal{B}_1$ is quartic in both variables, at first glance there is no obvious change of variables that allows us to rationalize the square root. In fact, homogenizing the polynomial as 
\begin{equation}
F_4(z_0,z_1,z_2) \equiv z_0^4 \, \mathcal{B}_1(z_1/z_0,z_2/z_0)\,,
\end{equation}
we find that the equation $y^2=F_4(z_0,z_1,z_2)$ defines a Del Pezzo surface of degree 2 in weighted projective space $[z_0:z_1:z_2:y] \in \mathbb{WP}^{1,1,1,2}$; cf.\ tab.~\ref{tab:geometries}. 
Del Pezzo surfaces are a special case of Fano varieties, and are known to be rationalizable~\cite{Schicho2005}. However, the Del Pezzo surface above is smooth, and it thus cannot be rationalized using the algorithm in refs.~\cite{Besier:2018jen,Besier:2019kco}.

Instead, we can use the rationalization procedure of ref.~\cite{Schicho2005}; see app.~\ref{app:rationalization_Del_Pezzo} for a step-by-step description.\footnote{We are grateful to Dino Festi for several enlightening discussions on ref.\ \cite{Schicho2005}.} Concretely, Theorem 24 of ref.~\cite{Schicho2005} guarantees the existence of a reparametrization of $[z_0:z_1:z_2:y]$ in terms of $[w_1 : w_2 : w_3:y]$, such that
\begin{align}
z_{n} = \! \sum\limits_{i,j,k=0}^{i+j+k \leq 3} \! \beta_{n;i,j,k} \, w_1^{i} w_2^{j} w_3^{k}\, , \qquad
y = \! \sum\limits_{i,j,k=0}^{i+j+k \leq 6} \! \gamma_{i,j,k} \, w_1^{i} w_2^{j} w_3^{k}\, ,
\end{align}
for $n=0,1,2$. Taking e.g.\ $w_3=1$ to dehomogenize, and introducing the Jacobian for the change of variables, we eventually obtain
\begin{equation}
\twoloop[210]{3}{3}{3} \, \Bigg|_{\text{max-cut}} \propto \int \frac{dw_1 dw_2}{P_6(w_1,w_2)} \, \det{\left[ \frac{\partial (z_1,z_2)}{\partial (w_1,w_2)} \right]} \,.
\label{eq:goombaP6}
\end{equation}
At this point, we could take a residue around a simple pole of the degree-6 polynomial $P_6(w_1,w_3)$. However, we could not find an analytic closed form for $P_6$ for generic kinematics, since it involves the solutions to polynomial equations of high degree; cf. app.~\ref{app:rationalization_Del_Pezzo}. Nevertheless, according to Theorem 24 of ref.~\cite{Schicho2005}, $P_6$ vanishes doubly at seven points, and its vanishing locus defines a curve in $\mathbb{P}^2$. For such curves, a genus-degree formula exists,
\begin{equation}
g = \frac{1}{2}(d-1)(d-2) - \frac{1}{2} \sum_s m_s (m_s-1)\,,
\end{equation}
where $d$ is the degree of the polynomial and the sum is over all singular points, with $m_s$ denoting their multiplicity. If the curve was smooth, i.e.\ if it had no singular points, it would generically have geometric (and arithmetic) genus 10. However, since $P_6$ vanishes doubly at seven points, it reduces the geometric genus to 3. As a consequence, the integral topologies $I_{3,3,3}$, $I_{3,3,3}^*$ and $I_{3,3,3}^{**}$ involve a generic curve of genus 3 at the maximal cut,\footnote{Specifically, we have numerically verified that the map from the space of kinematics to the 15 coefficients of the polynomial $\mathcal{B}_1$ has a Jacobian with full rank. This guarantees that the polynomial is generic, which carries over to the vanishing locus of $P_6(w_1,w_2)$ in eq.~\eqref{eq:goombaP6}.} which could in principle lie beyond the hyperelliptic realm.

Finally, even though neither $\mathcal{E}_1$ nor $\mathcal{B}_2$ contribute to the master integrals, let us investigate their associated geometry. First, since they are quadratic polynomials, they do not introduce a non-trivial geometry on their own. Then, we can consider the combination of $\mathcal{E}_1$ with $\mathcal{B}_1$. In this case, taking a residue at the point where $\mathcal{E}_1$ vanishes yields a square root of a polynomial of degree 8. This polynomial is, however, a perfect square. This degeneration is again explained by eq.~\eqref{eq: degeneration_Desnanot_Jacobi}, which follows from the Desnanot--Jacobi identity.

The inclusion of dots and numerators will not change the geometric nature of the leading singularities such that our results hold for all integrals in these sectors.

\subsection[Non-planar double heptagon]{\texorpdfstring{$\text{I}_{\text{4,3,4}}$}{I434} and starred versions: Non-planar double heptagon}
\label{sec:special_case_434}

Lastly, let us consider the integral topologies $I_{4,3,4}$, $I_{4,3,4}^*$ and $I_{4,3,4}^{**}$, corresponding to the non-planar double heptagon,
\begin{equation}
\twoloop[210]{4}{3}{4}
\end{equation}
These integral topologies are rather special, since their loop-by-loop Baikov representation does not simply follow from eq.~\eqref{eq:Baikov}. In particular, these are the only topologies from our analysis where both $a+b \geq 7$ and $b+c \geq 7$. Importantly, within the 't Hooft--Veltman scheme, this means that there are more propagators (in this case 7) than scalar products involving the loop momenta (in this case 6) for each loop, cf.\ sec.~\ref{sec: LBL_Baikov}. As a consequence, the propagators are not linearly independent when expressed in terms of the Baikov variables, and the loop-by-loop Baikov representation must be derived by other means, see e.g.\ refs.~\cite{Weinzierl:2022eaz, Frellesvig:2024zph, Frellesvig:2024ymq}.\ Concretely, we can calculate an induced Baikov representation using only a linearly independent subset of the propagators and multiply it by the remaining propagators, where we replace the scalar products by the respective Baikov variables. Alternatively, since the integral topologies $I_{4,3,4}$, $I_{4,3,4}^*$ and $I_{4,3,4}^{**}$ contain 11 propagators each, we can simply use the standard Baikov representation and match the 11 scalar products involving the loop momenta to the different propagators. Either way results in a parametrization without any ISPs, and the maximal cut for the corner integrals becomes algebraic, indicating that they admit a $d$log form on the maximal cut.

\section{Conclusions and outlook}
\label{sec:conclusions}

In this paper, we have classified the geometries occurring in all two-loop Feynman integrals for generic four-dimensional Quantum Field Theories with standard quadratic propagators, importantly including the Standard Model. Notably, these geometries determine the space of functions that appear in the result of the scattering amplitudes, and therefore in the physical observables. While we used the 't Hooft--Veltman scheme for our analysis, our results hold for the space of functions of any scheme-independent quantity, including observables. To achieve this classification, we have used a loop-by-loop Baikov parametrization to analyze the leading singularities for the integrals pertaining to a basis of 79 independent two-loop topologies. In addition, we have employed the notion of configuration matrices, which can be used to detect the potential presence of non-trivial geometries at the level of the maximal cut. Then, through non-trivial changes of variables and the rationalization of square roots, we have been able to explicitly express the leading singularities as integrals over non-trivial geometries in the pertinent cases.

Our results show that the most complicated geometries that occur at two-loop order are elliptic curves, K3 surfaces, hyperelliptic curves of genus 2 and 3, as well as a (smooth and non-degenerate) Del Pezzo surface of degree 2, which we could rationalize, revealing a curve of geometric genus 3; see figs.~\ref{fig: non-trivial_diagrams_2loop} and~\ref{fig: non-trivial_diagrams_2loop_evanescent} for an overview. To our knowledge, this is the first time that the rationalization of a square root associated to such a non-trivial Fano variety -- the Del Pezzo surface -- has occurred in the calculation of Feynman integrals. As a further result, we observe that elliptic curves are the most complicated geometries appearing in planar two-loop Feynman integrals. Conversely, for 2- and 3-point processes (including non-planar diagrams) elliptic curves are the most intricate geometries, while for 4-point processes K3 surfaces and genus-2 curves also occur. Lastly, for 5-point processes genus-3 curves appear, and for 6-point (and higher-point) processes Del Pezzo surfaces arise too.

Note that in this paper we have analyzed the geometries of Feynman integrals at the level of the maximal cut. Away from the maximal cut, a Feynman integral not only contains these geometries, but also the geometries of all of its subsectors, which we have equally classified. For instance, most two-loop integrals with generic masses contain different versions of the elliptic sunrise as a subsector; thus, they inherit its ellipticity.

Our results lay the foundation for evaluating the corresponding Feynman master integrals, for instance through the differential-equations method~\cite{Kotikov:1990kg}, which then can be used to compute the observables. In particular, the loop-by-loop Baikov representation we have used becomes a crucial ingredient for bringing the differential equation system into the so-called $\varepsilon$-factorized form~\cite{Henn:2013pwa,Gorges:2023zgv,Duhr:2025lbz,Maggio:2025jel,e-collaboration:2025frv,Duhr:2025xyy,Bree:2025tug,Becchetti:2025oyb}.

A first step towards evaluating all master integrals at two loops would be the calculation of all two-loop planar master integrals. In the planar case, we have shown that at most elliptic curves occur, for which the corresponding space of functions is increasingly well understood~\cite{Levin:2007tto,Brown:2011wfj,Broedel:2014vla,Broedel:2017kkb,Broedel:2018iwv,Broedel:2018qkq}. Along these lines, recent work in maximally supersymmetric Yang--Mills ($\mathcal{N}=4$ SYM) theory has successfully calculated the two-loop integrals forming the basis of planar scattering amplitudes~\cite{Morales:2022csr,Spiering:2024sea}. Together with the corresponding coefficients obtained via unitarity~\cite{Bourjaily:2015jna,Bourjaily:2017wjl}, this yields all planar two-loop amplitudes in that theory. Similarly for QCD and the Standard Model, modern methods can be applied to efficiently determine the coefficients of corresponding basis integrals, e.g.\ at 5-points~\cite{Ita:2015tya,Abreu:2020xvt,DeLaurentis:2023nss,DeLaurentis:2023izi,Agarwal:2023suw,Badger:2024fgb,Becchetti:2025qlu} with the use of pentagon functions~\cite{Papadopoulos:2015jft,Gehrmann:2018yef,Chicherin:2018old,Chicherin:2020oor,Chicherin:2021dyp,FebresCordero:2023pww,Badger:2024fgb}.

In this paper, we have considered two-loop integrals with generic values of the masses and off-shell external momenta. By contrast, only a limited number of different masses occur in the Standard Model, which will in many cases reduce the complexity of the associated Feynman-integral geometries. From our analysis, one can simply test whether the geometries degenerate for specific values of masses and momenta. For the elliptic and higher-genus curves, this can be easily done through the discriminant of the defining polynomials. For the K3 surface occurring in the tardigrade integral topologies, an analysis of the Picard rank and more severe degeneracies can be carried out using the tools of ref.~\cite{Doran:2023yzu}. The analysis of degeneracies for the Del Pezzo surface of degree 2 becomes however more intricate, and we leave it for future work, as it only starts to contribute to order $\mathcal{O}(\varepsilon^0)$ at three loops. In upcoming work~\cite{Bargiela:toapp}, we investigate the concrete geometries and special functions contributing to a number of different two-loop processes on the Les Houches wishlist~\cite{Andersen:2024czj,Huss:2025nlt}.

\section*{Acknowledgements}

We thank Dino Festi, Pierre Lairez, Pierre Vanhove and Stefan Weinzierl for fruitful discussions, as well as Andrew Harder, Eric Pichon-Pharabod and Duco van Straten for communication. 

P. Bargie\l{}a was supported by the Swiss National Science Foundation (SNF) under contract 200020-204200 and the European Research Council (ERC) under the European Union’s Horizon Europe research and innovation program grant agreement 101163627 (ERC Starting Grant ``AmpBoot''). P. Bargie\l{}a and R. Marzucca were supported by the European Research Council (ERC) under the European Union’s Horizon 2020 research and innovation program grant agreement 101019620 (ERC Advanced Grant TOPUP). R. Marzucca was further supported by the European Union’s Horizon 2020 research and innovation program EWMassHiggs (Marie Sk\l{}o-dowska Curie Grant agreement ID: 101027658) and the Emeritus Foundation. H. Frellesvig was supported in part by the Excellent Young Scientists Fund Program of the National Natural Science Foundation of China (NSFC). The work of R. Morales, F. Seefeld and M. Wilhelm was supported by the research grant 00025445 from Villum Fonden. R. Morales was also supported in part by Department of Energy grant DE-SC0007859 and the Leinweber Postdoctoral Fellowship from the University of Michigan. F. Seefeld and M. Wilhelm were further supported by the Sapere Aude: DFF-Starting Grant 4251-00029B. 

\appendix

\section{Non-trivial factorization of Picard--Fuchs operators}
\label{app:factorization_PF}

In this appendix, we exemplify the non-trivial factorization of Picard--Fuchs operators for multi-scale Feynman integrals with one particular case. Specifically, we show that even though the integral admits a $d$log form at the maximal cut, there is no rational factorization of the Picard--Fuchs operator, as its factorization necessarily involves square roots in the kinematic variables. Instead, we achieve a factorization by leveraging the results for the leading singularities of the integral.\footnote{A similar observation was made in refs.\ \cite{Badger:2024fgb,Becchetti:2025qlu}. In all these cases, the non-trivial factorization occurs together with nested square roots in the leading singularities. We thank Simone Zoia for this comment.}

Specifically, let us consider the corner integral in the topology $I_{4,1,3}^{*}$, also known as the 6-point penta-box. This integral generically depends on 8 internal masses and 14 external kinematic variables. To reduce it to a univariate problem, let us take an arbitrary kinematic line parametrized by $t$, i.e.\ we rescale all Mandelstam variables by $t$ and choose an arbitrary rational numerical point for all kinematic variables except for $t$. 
Then, we can consider the Picard--Fuchs operator of the integral with respect to the variable $t$~\cite{Lairez:2022zkj}, resulting in this case in a second-order operator $\mathcal{L}_2$. As explained in sec.~\ref{sec: identifying_geometries}, one way to determine the associated geometry is through the rational factorization of the Picard--Fuchs operator, which can be obtained e.g.\ via the \texttt{\textup{DFactor}} command implemented in \texttt{\textup{Maple}}. In this case, the operator $\mathcal{L}_2$ a priori does not factorize further, suggesting the presence of an elliptic curve. By contrast, the analysis of leading singularities for the corner integral yields two different algebraic results, compatible with a $d$log form on the maximal cut, indicating that the Picard--Fuchs operator should actually factorize into two (possibly different) first-order operators, $\mathcal{L}_2 = \mathcal{L}_1 \widetilde{\mathcal{L}}_1$.

To clarify the origin of this apparent discrepancy, let us look more closely at the leading singularity. The integral topology $I_{4,1,3}^{*}$ depends on one ISP, denoted as $z$, and the corner integral takes the following form on the maximal cut, where we drop all constant prefactors,
\begin{equation}\twoloop[1]{4}{1}{3} \, \Bigg|_{\text{max-cut}} \propto \int \frac{dz}{\sqrt{\mathcal{B}_1} \, \mathcal{B}_2} \, .
\end{equation}
Both Baikov polynomials $\mathcal{B}_1$ and $\mathcal{B}_2$ are quadratic polynomials in $z$, whose coefficients depend on $t$. Thus, we can use eq.~\eqref{eq: change_of_variables_(z-r1)(z-r2)} to rationalize the square root with respect to $z$ and subsequently take a residue at one of two poles. The resulting two residues differ with respect to their sign inside of a nested square root. Let us denote the corresponding two leading singularities as $\omega_1(t)$ and $\omega_2(t)$. Importantly, they are algebraic, and contain square roots that only depend on the kinematic variable $t$.

By construction, the Picard--Fuchs operator annihilates the corner integral on the maximal cut; cf.\ sec.~\ref{sec: identifying_geometries}. Therefore, it can also be constructed as the operator that manifestly annihilates both leading singularities $\omega_1$ and $\omega_2$; see e.g.\ ref.~\cite{Frellesvig:2024rea} for a discussion in the context of Feynman integrals. Concretely, we should have that $\mathcal{L}_2 = \mathcal{L}_1 \widetilde{\mathcal{L}}_1$, with
\begin{align}
\widetilde{\mathcal{L}}_1 =& \, \frac{\partial}{\partial t} - \frac{\omega_2'}{\omega_2} \, , \\
\mathcal{L}_1 =& \, \frac{\partial}{\partial t} - \frac{\frac{\partial}{\partial t} (\widetilde{\mathcal{L}}_1 \omega_1)}{\widetilde{\mathcal{L}}_1 \omega_1} \nonumber \\
= &\, \frac{\partial}{\partial t} - \frac{1}{\omega_1' \omega_2 - \omega_1 \omega_2'} \left( \omega_1'' \omega_2 - \omega_1 \omega_2'' - \omega_1'\omega_2' + \frac{\omega_1 (\omega_2')^2}{\omega_2} \right) \, ,
\end{align}
where we introduce the notation $\omega_i' \equiv \partial \omega_i / \partial t$ and $\omega_i'' \equiv \partial^2 \omega_i / \partial t^2$. As can be seen, $\widetilde{\mathcal{L}}_1$ by construction annihilates $\omega_2$, while $\mathcal{L}_1$ precisely annihilates the result of $\widetilde{\mathcal{L}}_1 \omega_1$. Constructed this way, the operator $\mathcal{L}_2 = \mathcal{L}_1 \widetilde{\mathcal{L}}_1$ manifestly annihilates both leading singularities $\omega_1$ and $\omega_2$. 

Expanding the product $\mathcal{L}_1 \widetilde{\mathcal{L}}_1$ reproduces exactly the same second-order operator $\mathcal{L}_2$ obtained through the differential-equations method, for which we could not find a factorization earlier on. This shows that the factorization procedure failed to detect $\mathcal{L}_1 \widetilde{\mathcal{L}}_1$ because these first-order operators contain square roots in the kinematic variable $t$; yet, the leading singularities are algebraic. Hence, with this counterexample, we have shown that computing the rational factorization of the Picard--Fuchs operator for multi-scale Feynman integrals is not sufficient to determine the underlying geometry.

\section{Gram determinants and the degree of Baikov polynomials}
\label{app:gram}

In this appendix, we derive general constraints on the degree of the Baikov polynomials in the Baikov variables, which are used in our analysis in sec.~\ref{sec: General_analysis_geometries}. 

Let us consider a Gram determinant involving $l$ different loop momenta and $e$ external momenta. We sort these momenta as
\begin{align}
\left\{ q \right\} = \left\{ k_1, \ldots, k_{l}, p_1, \ldots, p_{e} \right\} \, .
\end{align}
The $(l+e) \times (l+e)$ Gram matrix then becomes
\begin{align}
G &= \left( \begin{array}{cc}
A & B \\ B^{T\!\!} & C
\end{array} \right) ,
\end{align}
where we define
\begin{align}
\label{eq: app_A_scalar_prods_Gram}
A_{ij}^{(l \times l)} = k_i \cdot k_j \,, \qquad B_{ij}^{(l \times e)} = k_i \cdot p_j \,, \qquad C_{ij}^{(e \times e)} = p_i \cdot p_j \,.
\end{align}

In the Baikov representation, cf.\ eq.~\eqref{eq:Baikov}, the Baikov polynomials depend on determinants of Gram matrices; see eqs.~\eqref{eq:Bdef_1}--\eqref{eq:Bdef_2}. However, to transform to the Baikov variables, we perform a linear shift from the scalar products in eq.~\eqref{eq: app_A_scalar_prods_Gram} to the Baikov variables $\vec{z}$, which also involves the Mandelstam variables and masses, which we collectively denote as $\vec{s}$ in the following. Therefore, the elements of $A$ and $B$ will depend on both $\vec{z}$ and $\vec{s}$ through functions we denote as $f_{ij}(\vec{z}, \vec{s})$, whereas $C$ will only depend on $\vec{s}$ through functions called $g_{ij}(\vec{s})$. We note that the functional dependence of $f_{ij}$ on $\vec{z}$ is linear. We thus have
\begin{align}
A_{ij}^{(l \times l)} = f_{ij}(\vec{z}, \vec{s}) \,, \qquad B_{ij}^{(l \times e)} = f_{i (j{+}l)}(\vec{z}, \vec{s}) \,, \qquad C_{ij}^{(e \times e)} = g_{ij}(\vec{s}) \,. 
\end{align}

Computing now the determinant of $G$, each term is given by a product of entries, such that exactly one entry comes from each row and (simultaneously) each column. Concretely, we get from the first $l$ rows (corresponding to $A$ and $B$) $l$ factors of $f$. In addition, from the next $l$ rows (corresponding to $B^{T}$ and $C$), we can get at most $l$ factors of $f$ on top. Since the functions $f$ are linear in the Baikov variables $\vec{z}$, each term in the determinant contains at most $2l$ powers of $\vec{z}$. As a consequence, any Gram determinant $\det G$ is at most of degree $2l$ in the Baikov variables.

When using the loop-by-loop Baikov representation for an $L$-loop integral, we thus find that $\mathcal{B}_1$ has at most degree $2L$, $\mathcal{E}_1$ and $\mathcal{B}_2$ have at most degree $2L-2$, etc., all the way down to $\mathcal{E}_L$, which does not depend on the Baikov variables. 

For the loop-by-loop Baikov representation at two loops, we obtain an extra constraint for planar integrals. In that case, we can split the scalar products involving loop momenta into three categories. The first type contains $k_1$ but not $k_2$, the second type contains $k_2$ but not $k_1$, and the third type (of which there is only one for a planar integral) will be $k_1 \cdot k_2$.\footnote{While this is not the case for all parametrizations of the loop momenta, we can always find a parametrization where it is the case.} The first type will only appear in the first row and column of $G$, the second type only in the second row and column of $G$, and the third type only in $A$, which is $2 \times 2$. Thus, for each of the three types, the determinant can have at most degree $2$. Therefore, we can deduce that while $\mathcal{B}_1$ will be of (at most) total degree $4$, it will be at most degree $2$ in the individual Baikov variables.

For the integral topologies $I_{a,1,1}$ with any number of stars, i.e.\ those that contain a bubble as one of its loops, $\mathcal{B}_1$ is the determinant of a two-by-two matrix since $e=0$, and thus it can at most be degree $2$ despite the discussion above.

\section{Rationalization of a Del Pezzo surface of degree 2}
\label{app:rationalization_Del_Pezzo}

In this appendix, we outline the rationalization procedure for a Del Pezzo surface of degree 2; see Theorem 24 of ref.~\cite{Schicho2005}. In particular, this variable transformation is used in sec.~\ref{sec:special_case_goomba} to rationalize the square root
\begin{equation}
\label{eq: app_initial_integral}
\int \frac{dz_1 dz_2}{\sqrt{\mathcal{B}_1(z_1,z_2)}}
\end{equation}
that appears after taking the maximal cut of the corner integral for the topology $I_{3,3,3}$, where $\mathcal{B}_1(z_1,z_2)$ is a polynomial of overall degree 4 and quartic in both variables.

Homogenizing the polynomial $\mathcal{B}_1$ as 
\begin{equation}
\label{eq: app_def_F4}
F_4(z_0,z_1,z_2) \equiv z_0^4 \, \mathcal{B}_1(z_1/z_0,z_2/z_0)
\end{equation}
in projective space $[z_0:z_1:z_2] \in \mathbb{P}^2$, the equation $y^2=F_4(z_0,z_1,z_2)$ defines a Del Pezzo surface of degree 2 in weighted projective space $[z_0:z_1:z_2:y] \in \mathbb{WP}^{1,1,1,2}$. 
\begin{figure}[t]
\begin{center}
\includegraphics[height=8.5cm]{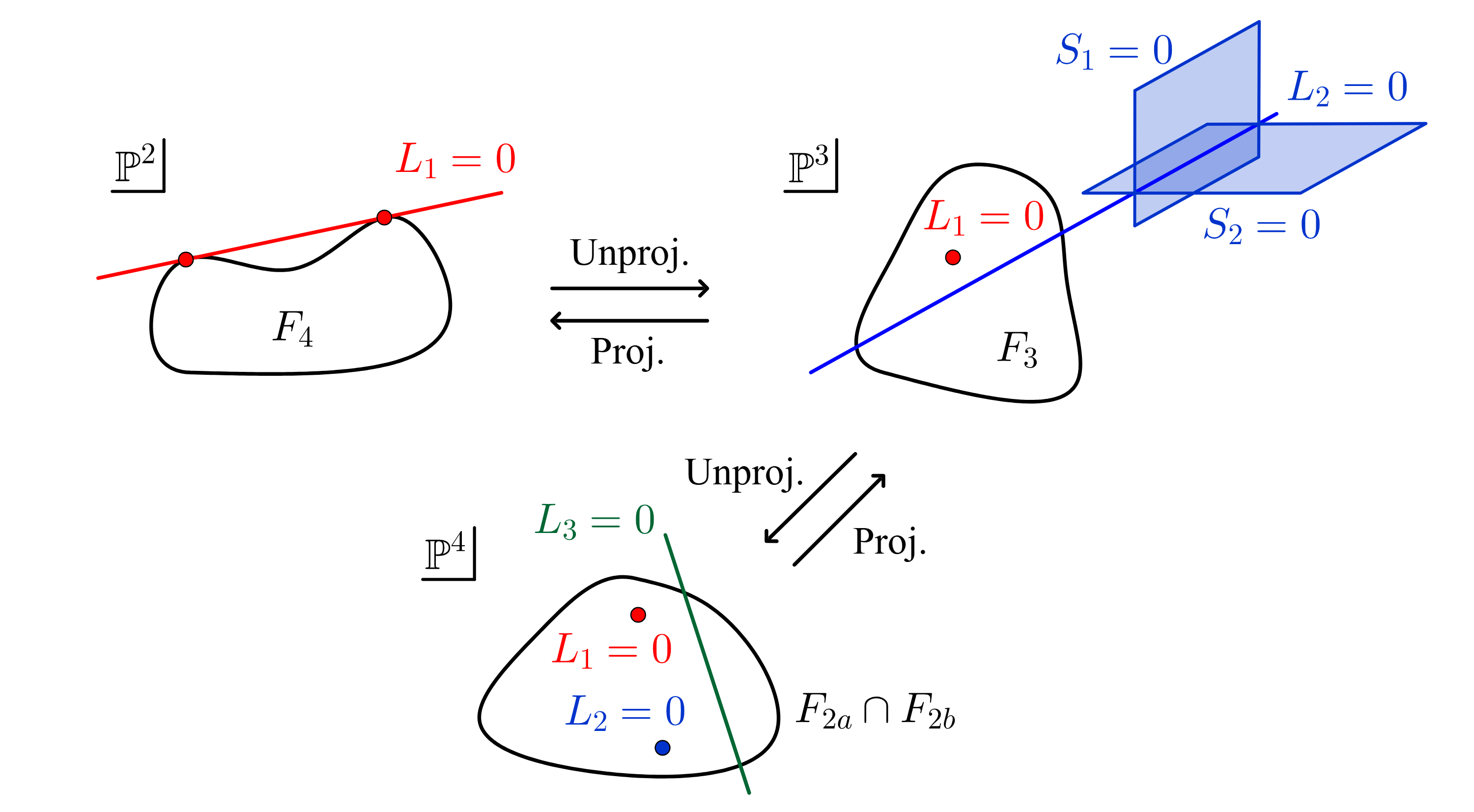}
\caption{Visual representation of the rationalization procedure outlined in this section, including the different surfaces and lines described, as well as the sequential unprojections from $\mathbb{P}^2$ up to $\mathbb{P}^4$ and the opposite projections.}
\label{fig: rationalization_Del_Pezzo}
\end{center}
\end{figure}
The aim of the rationalization is to find a change of variables such that $F_4(z_0,z_1,z_2)$ becomes a perfect square. With this goal, the starting point is to find a bitangent line to $F_4$, i.e.\ a line that is tangent to $F_4$ at two distinct points; see fig.~\ref{fig: rationalization_Del_Pezzo} for a visual representation. 
Any smooth quartic such as $F_4$ has exactly 28 bitangents~\cite{Schicho2005}.
Concretely, we are looking for a linear form
\begin{equation}
L_1(z_0,z_1,z_2) \equiv \alpha_0 \, z_0 + \alpha_1 \, z_1 + \alpha_2 \, z_2 \, ,
\end{equation}
such that $F_4$ restricted to the line $L_1 = 0$ vanishes doubly at two different points. In practice, we first solve $L_1=0$, e.g.\ with 
\begin{equation}
[z_0 : z_1 : z_2 ] = [ - (\alpha_1 \, z_1 + \alpha_2)/\alpha_0 : z_1 : 1] \, ,
\end{equation}
where we dehomogenize by setting $z_2=1$. Then, we impose that
\begin{equation}
\label{eq: app_F4_restricted_line}
F_4 \Big|_{L_1=0} = F_4 (- (\alpha_1 \, z_1 + \alpha_2)/\alpha_0 , z_1 , 1) = A \, (z_1-r_1)^2 (z_1-r_2)^2 \, ,
\end{equation}
where $A,r_1,r_2$ are constants, with $r_1 \neq r_2$. Thus, we can solve for $A,r_1,r_2$ and $[\alpha_0 : \alpha_1 : \alpha_2 ] \in \mathbb{P}^2$ order-by-order in $z_1$ in eq.~\eqref{eq: app_F4_restricted_line}, resulting in five equations of degree 4 and five unknowns, where we dehomogenize by setting $\alpha_0=1$ at the end. Thereafter, we can define
\begin{equation}
Q(z_1,z_2)^2 \equiv F_4 (- \alpha_1 \, z_1 - \alpha_2 z_2 , z_1 , z_2) \big|_{\text{Sol}(\alpha_1,\alpha_2)} \, ,
\end{equation}
where we substitute one of the previous solutions. This way, we have
\begin{equation}
F_4(z_0,z_1,z_2) = L_1 (z_0,z_1,z_2) \, P_3(z_0,z_1,z_2) + Q(z_1,z_2)^2 \,,
\end{equation}
where $P_3(z_0,z_1,z_2)$ is a cubic polynomial. Thus, on the line $L_1=0$, we automatically fulfill eq.~\eqref{eq: app_F4_restricted_line}, and $F_4$ becomes a perfect square.

The next step in the procedure of ref.\ \cite{Schicho2005} is to unproject from $[z_0:z_1:z_2] \in \mathbb{P}^2$ onto $[y_0:y_1:y_2:y_3] \in \mathbb{P}^3$ by mapping $z_0 \to y_0$, $z_1 \to y_1$ and $z_2 \to y_2$, and introducing a new variable $y_3$. Then, let us define a cubic form
\begin{equation}
F_3(y_0,\dots,y_3) \equiv \frac{1}{L_1} \left( F_4 - (L_1 \, y_3 + Q)^2 \right)\,,
\end{equation}
where we drop the dependence on $y_0,y_1,y_2$ on the right-hand side for ease of notation. Now, we seek for a line $L_2=0$ on $\mathbb{P}^3$, such that $F_3$ restricted to the line vanishes, which again realizes that $F_4$ is a perfect square. In particular, we can obtain such a line from the intersection of two planes $S_1=0$ and $S_2=0$; see fig.~\ref{fig: rationalization_Del_Pezzo}. Without loss of generality, we choose the planes to be perpendicular and defined via
\begin{align}
S_1(y_0,\dots,y_3) \equiv & \, y_0 + \beta_1 y_1 + \beta_2 y_2 \,,\\
S_2(y_0,\dots,y_3) \equiv & \, - (\beta_1 \beta_3+\beta_2 \beta_4) y_0 + \beta_3 y_1 + \beta_4 y_2 + y_3 \,.
\end{align}
To obtain the line $L_2=0$, we solve for
\begin{equation}
F_3 (y_0,\dots,y_3) = S_1 \, P_2 + S_2 \, \widetilde{P}_2 \,,
\end{equation}
where $P_2$ and $\widetilde{P}_2$ are quadratic polynomials in $y_0,\dots,y_3$. This way, the line $L_2 = S_1 \cap S_2 = 0$ automatically fulfills that $F_3=0$, and as a byproduct $F_4=(L_1 y_3 + Q)^2$ becomes a perfect square.

Next, we unproject again, onto $[x_0:x_1:x_2:x_3:x_4] \in \mathbb{P}^4$ by mapping $y_0 \to x_0$, \dots, $y_3 \to x_3$, and introducing a fourth variable $x_4$; see fig.~\ref{fig: rationalization_Del_Pezzo} for reference. Then, let us define two quadratic forms
\begin{align}
F_{2a}(x_0,\dots,x_4) \equiv & \, P_2 + x_4 S_2 \,,\\
F_{2b}(x_0,\dots,x_4) \equiv & \, - \widetilde{P}_2 + x_4 S_1 \,.
\end{align}
Once again, we seek for a line $L_3=0$ such that $F_{2a}=F_{2b}=0$, which in turn implies that $F_3=0$ and that $F_4$ is a perfect square. In this case, the line can be obtained through the intersection of three hypersurfaces $H_1=0$, $H_2=0$ and $H_3=0$, which we take as
\begin{align}
H_1(x_0,\dots,x_4) \equiv & \, x_0 + \alpha_{1,1} \, x_1 + \alpha_{1,2} \, x_2 \,,\\
H_2(x_0,\dots,x_4) \equiv & \, x_3 + \alpha_{2,1} \, x_1 + \alpha_{2,2} \, x_2 \,,\\
H_3(x_0,\dots,x_4) \equiv & \, x_4 + \alpha_{3,1} \, x_1 + \alpha_{3,2} \, x_2 \,.
\end{align}
Thus, we can obtain the line $L_3=0$ by solving
\begin{equation}
F_{2a}\big|_{H_1=H_2=H_3=0} = F_{2b}\big|_{H_1=H_2=H_3=0} = 0
\end{equation}
for $\alpha_{i,j}$. 

Finally, to obtain the rationalization of the Del Pezzo surface, we need to find all planes passing through the line $L_3=0$. This defines a map from $\mathbb{P}^4$ onto $\mathbb{P}^2$. In other words, we can define
\begin{equation}
p_1 \equiv F_{2a}\big|_{H_i = w_i} \, , \quad p_2 \equiv F_{2b}\big|_{H_i = w_i} \, ,
\end{equation}
with $[w_1:w_2:w_3] \in \mathbb{P}^2$ expressed in terms of $[x_0:x_1:x_2:x_3:x_4] \in \mathbb{P}^4$, and invert the map by solving the equations
\begin{equation}
p_1 \big|_{\text{Sol}(\alpha_{i,j})} =0 \, , \quad p_2 \big|_{\text{Sol}(\alpha_{i,j})} =0 \, ,
\end{equation}
where we substitute the previous solution for $\alpha_{i,j}$. Concretely, solving these equations we obtain a cubic change of variables for $x_i$ in terms of $w_i$, so that $F_4$ is a perfect square. Then, the last step consists in projecting back from $[x_0:x_1:x_2:x_3:x_4] \in \mathbb{P}^4$ to $[z_0:z_1:z_2] \in \mathbb{P}^2$ by mapping $x_0 \to z_0$, $x_1 \to z_1$, $x_2 \to z_2$ and substituting the cubic change of variables in eq.~\eqref{eq: app_def_F4}; cf.\ fig.~\ref{fig: rationalization_Del_Pezzo}. 

At the end, $F_4$ becomes a perfect square of a degree-6 polynomial $P_6(w_1,w_2,w_3)$ that vanishes doubly at 7 points~\cite{Schicho2005}. The final step is to dehomogenize by setting $w_3=1$, after which eq.~\eqref{eq: app_initial_integral} becomes
\begin{equation}
\int \frac{dw_1 dw_2}{P_6(w_1,w_2)} \, \det{\left[ \frac{\partial (z_1,z_2)}{\partial (w_1,w_2)} \right]} \,,
\end{equation}
where we introduce the Jacobian of the transformation.

Lastly, let us note that this rationalization procedure involves solving polynomial equations of degree as high as 9, and subsequently using their roots in other equations. Therefore, we are not able to find an analytic closed form for the degree-6 polynomial $P_6(w_1,w_2)$. In practice, most steps in the rationalization are performed numerically, but we leave a study of the analytic structure of $P_6$ for future work.

\bibliography{References}
\bibliographystyle{JHEP}

\end{document}